\documentclass[twocolumn]{aastex631}

\usepackage{booktabs} 
\usepackage{xcolor}
\usepackage{amsmath,bm}
\usepackage{CJKutf8}
\usepackage{hyperref}
\usepackage{tabularx}
\usepackage{fancyhdr}
\usepackage{longtable} 
\usepackage{float}

\newcommand{\frb}{FRB 20190520B}

\newcommand{\AppFig}[1]{Appendix Figure~\ref{#1}}

\newif\ifshowrevisions
\showrevisionsfalse 

\DeclareRobustCommand{\rev}[1]{%
  \ifshowrevisions \textcolor{blue}{\textbf{#1}}%
  \else #1\fi
}

\newenvironment{revblock}
  {\ifshowrevisions\color{blue}\bfseries\boldmath\fi}
  {}

\begin{document}

\title{Statistical and Temporal Analysis of Multi-component Burst-clusters \\ from the Repeating FRB 20190520B}
\shorttitle{multi-component bursts of \frb}
\shortauthors{zhang et al.}

\author[0009-0009-3949-4726]{Jia-heng Zhang}
\affiliation{Institute of Astrophysics, Central China Normal University, Wuhan 430079, China}
\affiliation{Education Research and Application Center, National Astronomical Data Center, Wuhan 430079, China}
\affiliation{Key Laboratory of Quark and Lepton Physics (Central China Normal University), Ministry of Education, Wuhan 430079, China}

\author[0000-0001-6651-7799]{Chen-Hui Niu*}
\affiliation{Institute of Astrophysics, Central China Normal University, Wuhan 430079, China}
\affiliation{Education Research and Application Center, National Astronomical Data Center, Wuhan 430079, China}
\affiliation{Key Laboratory of Quark and Lepton Physics (Central China Normal University), Ministry of Education, Wuhan 430079, China}
\email{niuchenhui@ccnu.edu.cn}

\author[0009-0009-8320-1484]{Yu-hao Zhu}
\affiliation{National Astronomical Observatories, Chinese Academy of Sciences, Beijing 100101, China}
\affiliation{School of Astronomy and Space Science, University of Chinese Academy of Sciences, Beijing 100049, China}

\author[0000-0003-3010-7661]{Di Li}
\affiliation{Department of Astronomy, Tsinghua University, Beijing 100084, China}
\affiliation{National Astronomical Observatories, Chinese Academy of Sciences, Beijing 100101, China}
\affiliation{Research Center for Astronomical Computing, Zhejiang Laboratory, Hangzhou 311100, China}

\author{Yu Wang}
\affiliation{ICRANet, Piazza della Repubblica 10, 65122 Pescara, Italy}

\author[0000-0001-9036-8543]{Wei-yang Wang}
\affiliation{School of Astronomy and Space Science, University of Chinese Academy of Sciences, Beijing 100049, China}

\author{Yi Feng}
\affiliation{Research Center for Astronomical Computing, Zhejiang Laboratory, Hangzhou 311100, China}
\affiliation{Institute for Astronomy, School of Physics, Zhejiang University, Hangzhou 310027, China}

\author[0009-0008-0639-3964]{Xin-ming Li}
\affiliation{Institute of Astrophysics, Central China Normal University, Wuhan 430079, China}
\affiliation{Education Research and Application Center, National Astronomical Data Center, Wuhan 430079, China}
\affiliation{Key Laboratory of Quark and Lepton Physics (Central China Normal University), Ministry of Education, Wuhan 430079, China}

\author[0000-0001-8065-4191]{Jia-rui Niu}
\affiliation{National Astronomical Observatories, Chinese Academy of Sciences, Beijing 100101, China}

\author[0000-0002-3386-7159]{Pei Wang}
\affiliation{National Astronomical Observatories, Chinese Academy of Sciences, Beijing 100101, China}

\author[0000-0002-1067-1911]{Yun-wei Yu}
\affiliation{Institute of Astrophysics, Central China Normal University, Wuhan 430079, China}
\affiliation{Education Research and Application Center, National Astronomical Data Center, Wuhan 430079, China}
\affiliation{Key Laboratory of Quark and Lepton Physics (Central China Normal University), Ministry of Education, Wuhan 430079, China}

\author[0000-0002-8744-3546]{Yong-kun Zhang}
\affiliation{National Astronomical Observatories, Chinese Academy of Sciences, Beijing 100101, China}

\author{Xiao-ping Zheng}
\affiliation{Institute of Astrophysics, Central China Normal University, Wuhan 430079, China}
\affiliation{Education Research and Application Center, National Astronomical Data Center, Wuhan 430079, China}
\affiliation{Key Laboratory of Quark and Lepton Physics (Central China Normal University), Ministry of Education, Wuhan 430079, China}

\begin{abstract}
Fast Radio Bursts (FRBs) are bright, millisecond-duration extragalactic radio transients that probe extreme astrophysical environments. Many FRBs exhibit multi-component structures, which encode information about their emission mechanisms or progenitor systems and thus provide important clues to their origins. In this work, we systematically analyze the burst morphology of FRB 20190520B and compare component distributions across four active FRBs observed with FAST: FRB 20121102A, FRB 20190520B, FRB 20201124A, and FRB 20240114A. We find that multi-component burst-clusters show spectral properties similar to single-peak bursts, and no periodicity is detected in their temporal behavior. The component-count distributions follow a power law, revealing scale-free behavior consistent with self-organized criticality (SOC) processes. Multi-component clusters account for $12$–$30\%$ of all detected bursts, regardless of source activity, providing new insights into burst-to-burst variability and the physical processes driving FRB emission.
\end{abstract}


\section{Introduction} \label{sec:intro}
Fast Radio Bursts (FRBs) are high-energy astrophysical phenomena characterized by short, intense bursts of radio emission lasting only milliseconds. Since their discovery in 2007~\citep{Lorimer_2007}, FRBs have posed significant challenges to our understanding of the astrophysics of compact objects. Many of these bursts exhibit high dispersion measures (DMs) that exceed the predictions of Galactic electron density models along their lines of sight~\citep{2004ne2001,2017YaoJM}, suggesting an extragalactic origin~\citep{2019cordes,2020Naturezhang, 2022Petroff}. A major breakthrough came with the identification of the first repeating source, FRB 20121102A, in 2016~\citep{2016Naturspilter}, which opened up unprecedented opportunities for follow-up studies. To date, over 800 FRB sources have been discovered, with roughly 10\% of them exhibiting repeating behavior\footnote{\href{https://blinkverse.zero2x.org}{https://blinkverse.zero2x.org}}.
 A subset of these have also been localized to host galaxies. In contrast to one-off FRBs, repeating FRBs offer a unique window into the underlying physical mechanisms and local environments of FRB sources~\citep{2019CHIME, 2020Fonseca, 2021chimecat}.

\begin{figure*}[t]
    \centering
    \includegraphics[width=0.8\linewidth]{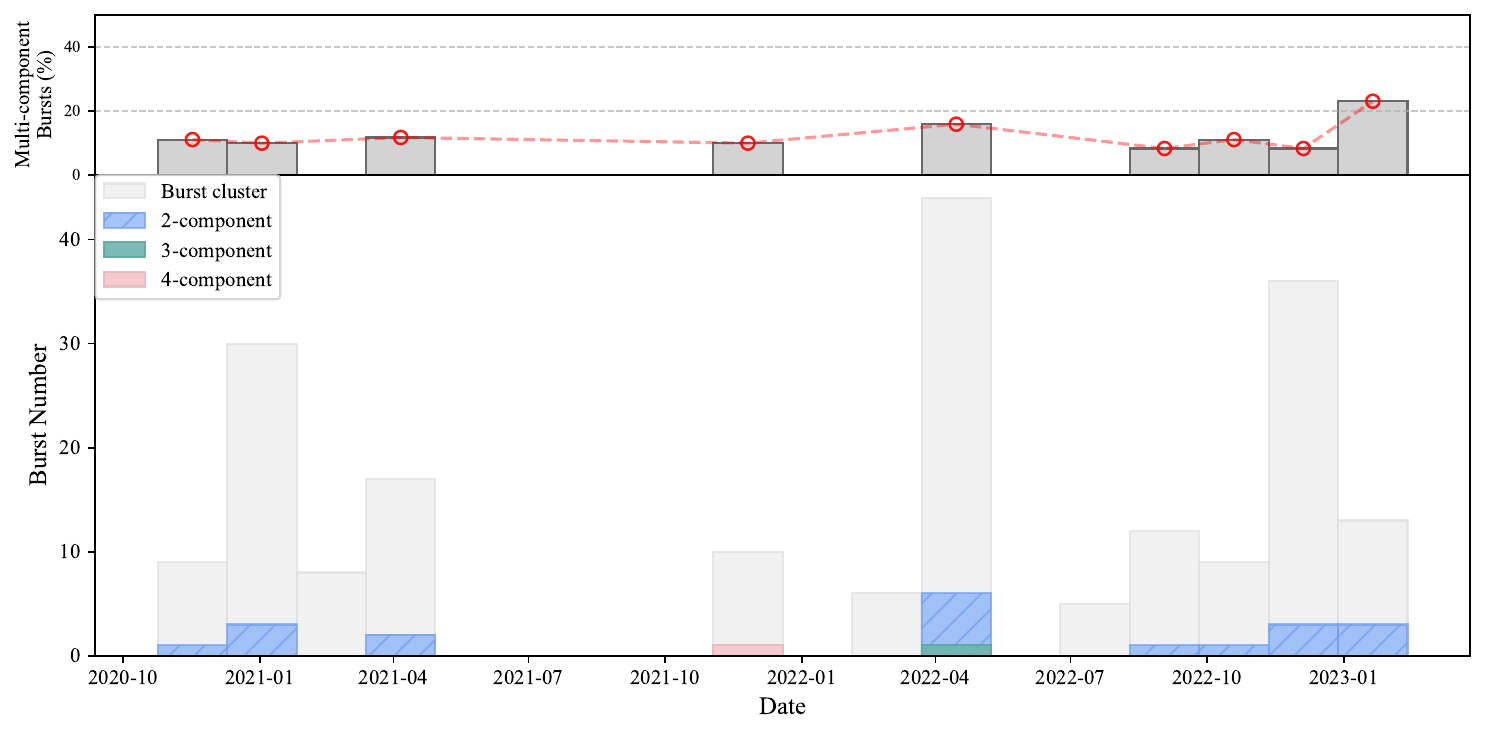}
    \caption{The horizontal axis represents the dates, and the vertical axis shows the number of bursts. Different colors are used to distinguish the total burst count, bimodal burst structure, trimodal burst structure, and quadrimodal burst structure.}
    \label{fig:Burst_number}
\end{figure*}

Multi-component structures in radio emission were firstly discovered in pulsars~\citep{1968craftpulsar}, and were subsequently analyzed in FRB~20121102A~\citep{2019Hessels}. Since then, such features have been observed in multiple FRB sources, characterized by complex multi-component structures and quasi-periodic patterns~\citep{zhou_2022, Pasto_2023, 2022NJR, 2021ApJmajid, 2024kramer}. The presence of these multi-component structures in FRBs has drawn a close analogy between FRBs and pulsars. Moreover, the temporal and spectral evolution of these fine structures may offer further insights into the underlying radiation mechanisms.

The periodicity of FRB multi-component structure has also been studied. In FRB 20201124A, a broad distribution of sub-burst separations around 5~ms was reported~\citep{2022NJR}, but without a consistent or dominant period, suggesting an origin in intrinsic radiation processes rather than rotational modulation. A tentative 1.7-second periodicity was observed in FRB 20201124A on two isolated days, potentially linked to magnetar rotation\citep{2025Duchen}, though its sporadic appearance leaves the origin uncertain.
Similar quasi-periodic features have also been observed across a range of timescales. FRB~20201020A exhibits a quasi-periodic structure with five subcomponents regularly spaced by 0.411~ms, suggesting a potential magnetospheric origin for such periodicity \citep{Pasto_2023}. \citet{2024kramer} reported that the quasi-periodic sub-structures seen in pulsar and magnetar radio emission scale with their rotational periods, suggesting a possible link between multi-component structures timescales and the spin of the neutron star. 
If FRBs originate from similar systems, such scaling could in principle be used to infer their underlying periods. However, no direct observational evidence has confirmed \rev{the validity of this relationship} in FRBs to date.

One of the most intriguing repeating sources, \frb, was discovered by the Five-hundred-meter Aperture Spherical Radio Telescope (FAST) and is co-located with a compact, luminous persistent radio source (PRS) in a star-forming dwarf galaxy at redshift z = 0.241 \citep{niu_2022}. This source exhibits a high repetition rate, complex burst morphology with multi-component structures, and an exceptionally large DM of $1204.7 \pm 4.0\,\mathrm{pc\,cm^{-3}}$, of which $\sim 903^{+72}_{-111}\,\mathrm{pc\,cm^{-3}}$ is attributed to the host galaxy. This host DM far exceeds that of other known FRBs and deviates significantly from the Macquart relation~\citep{2020macquart}, implying that a substantial fraction of the dispersion originates in its local environment rather than in the intergalactic medium. The persistent radio counterpart has a luminosity of $\sim 3 \times 10^{29}\,\mathrm{erg\,s^{-1}\,Hz^{-1}}$, and its compactness and brightness closely resemble those of FRB~20121102A. These similarities suggest that \frb \ and FRB~20121102A may represent a distinct subclass of young repeating FRBs embedded in dense magneto-ionic environments, making FRB~20190520B an ideal laboratory for studying the interplay between FRB emission and its local environment.

\begin{figure}[htbp]
    \centering
    \includegraphics[width=0.5\textwidth]{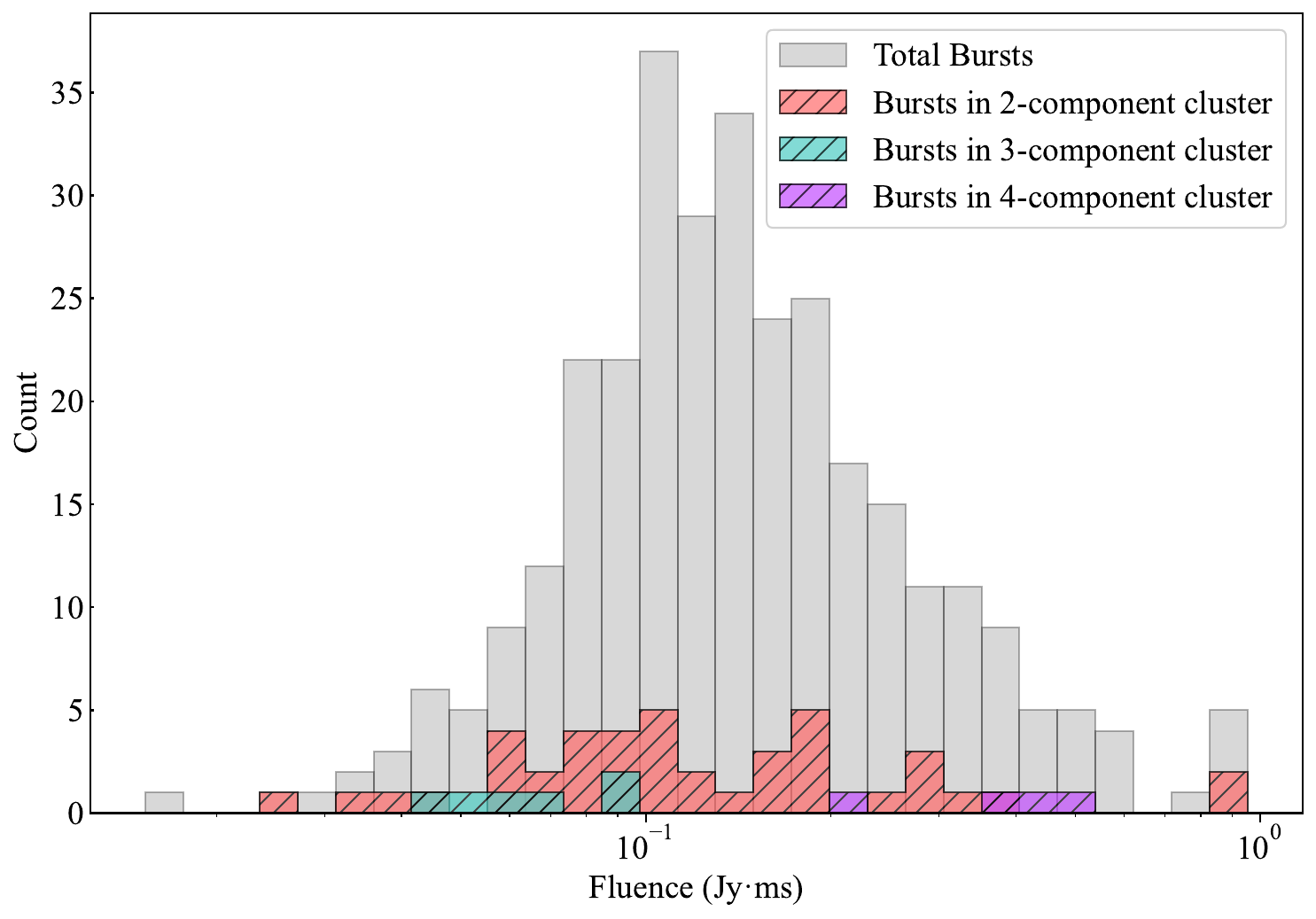}
    \caption{
    Distribution of burst fluences detected from \frb. 
    The horizontal axis represents the \rev{fluence (Jy$\cdot$ms)}, and the vertical axis shows the burst count.
    The gray bars denote the total distribution of all bursts, while the hatched colored bars highlight bursts that belong to multi-component structures: 
    red for bursts in 2-component clusters, cyan for bursts in 3-component clusters, and purple for bursts in 4-component clusters.
    }
    \label{fig:fluence_distribution}
\end{figure}

In this paper, we focus on the overall morphological characteristics of multi-component FRB bursts of \frb, analyzing their time-frequency behavior, spectral distribution, and the statistical properties of component counts. We choose \frb\ as our target because its sustained activity allows continuous monitoring on yearly timescales, and its bursts often exhibit rich multi-component structures that are ideal for statistical investigation. By statistically examining a large sample of bursts from a repeating source, we aim to provide a refined understanding of the emission structure and its physical implications. Sections~\ref{Observations} and~\ref{DATA} describe our observations, burst detection, and data reduction procedures, including flux calibration. Section~\ref{Analysis} presents the analysis of multi-component burst-clusters, focusing on their time–frequency behavior, spectral features, and temporal properties. Section~\ref{Result} summarizes the main results, compares the component count statistics of \frb\ with those of other repeating FRB sources. Section~\ref{sec:Dissusion} provides a discussion of the results and the overall conclusions.

\section{Observations}
\label{Observations}
\frb\ was observed with the Five-hundred-meter Aperture Spherical Radio Telescope (FAST), using the central beam of the 19-beam L-band receiver and the pulsar backend. The system covered the frequency range 1.0–1.5 GHz (centered at 1.25 GHz) with a time resolution of 49.152~$\mu$s and a frequency resolution of 0.122~MHz. \rev{The data were recorded in the standard \textsc{psrfits} with full-Stokes polarization information.}

The observations and search procedures are identical to those described in \citet{niu_2022} and \rev{\citet{Niu2025_inprep}, }and we summarize them here for completeness. Candidate bursts were identified with the \textsc{Heimdall} pipeline \citep{heimdall}, applying a zero-DM matched filter for RFI excision. Events with ${\rm S/N} > 7$ were visually inspected using their dynamic spectra to confirm astrophysical origin.

Flux calibration was carried out following the standard procedure \rev{in\citet{Niu2025_inprep}, }using an injected periodic noise signal with a known reference temperature before each observation. From the calibrated data, we derived the burst properties including DM, temporal width, center frequency, bandwidth, peak flux density, fluence, and isotropic-equivalent energy \citep{ZhangB2018, Gourdji2019}.

From October 24, 2020, to April 1, 2023, a total of 72 observing sessions were conducted, corresponding to 72.04 hours of on-source time. In this new \rev{search~\citep{Niu2025_inprep}}, 360 bursts were detected. The burst parameters were uniformly measured and compiled into a catalog, which serves as the basis for the statistical analyses presented in the following sections. The average event rate across all sessions was approximately 5.0 bursts per hour.

\begin{figure}[b]
    \centering
    \includegraphics[width=0.5\textwidth]{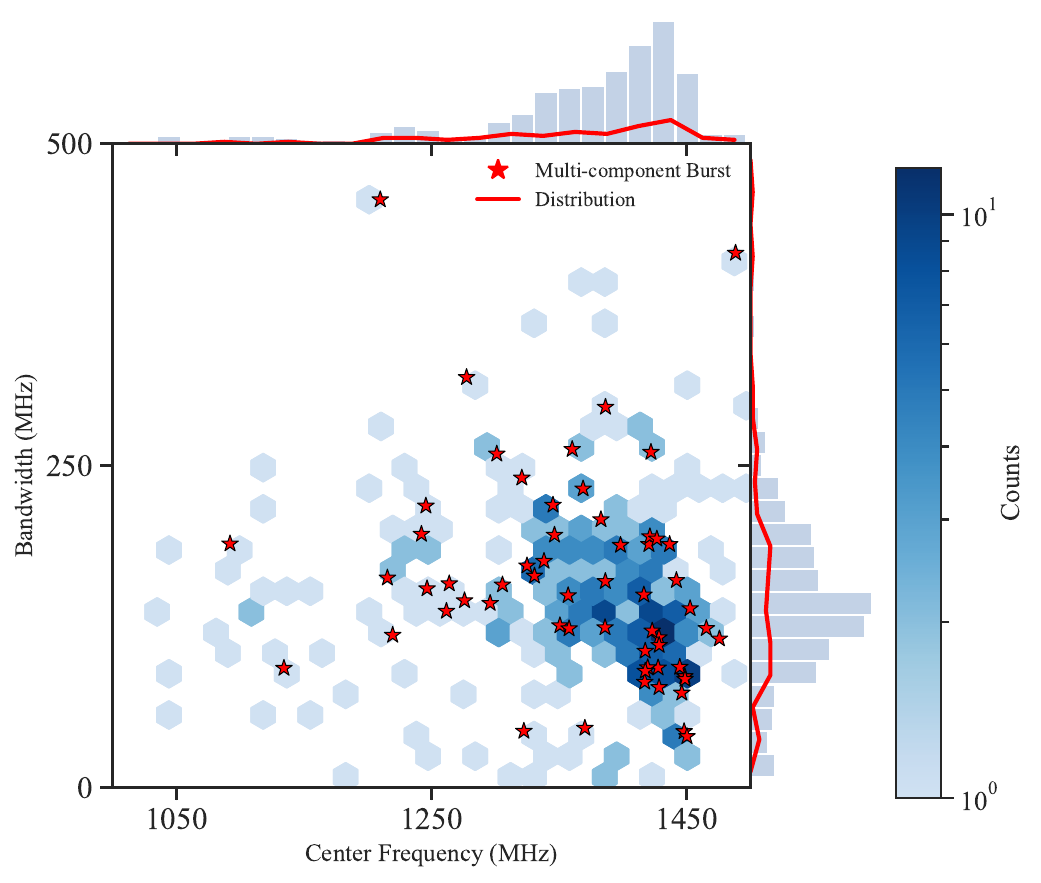}
    \caption{
    Distribution of burst center frequencies and bandwidths for FRB 20190520B. 
    The blue hexagons represent all detected bursts, with the color intensity indicating the number of bursts in each bin. 
    Red stars mark bursts that belong to multi-component burst-clusters. 
    The marginal histograms along the top and right axes show the overall distributions of burst center frequency and bandwidth, respectively. 
    The red curve represents the distribution of the multi-component burst-clusters (red stars) in center frequency and bandwidth.
    }
    \label{fig:freq_bw_distribution}
\end{figure}    

\section{DATA}
\label{DATA}
In this section, we present the criteria used to identify burst-clusters in our dataset and describe the characteristics of the bursts within them. Following \citet{zhou_2022} and \citet{2025ZhangLX}, we define a \textit{burst-cluster} as a collection of bursts (excluding those labeled as ``Not-Clear'') in which the time separations between adjacent components (or peaks) are shorter than a threshold $\tau$. The threshold $\tau$ is determined from the valley in the bimodal waiting-time distribution; for example, \citet{2025ZhangJS} derived $\tau = 400$~ms for FRB~20240114A (see their Extended Data Figure~2). A single burst is still classified as a burst-cluster if no other bursts are detected within the $\tau$-duration intervals immediately preceding and following it.

While \citet{zhou_2022} and \citet{2025ZhangLX} further distinguish between \textit{sub-bursts} (quasi-connected components with visually separable peaks in the de-dispersed profile) and \textit{bursts} (groups of such sub-bursts), in this work we do not make this distinction. For convenience in our multi-component structure analysis, we refer to any individual component within a burst-cluster as a \textit{burst}, provided that the difference between its peak and the adjacent valley exceeds $3\sigma$ above the baseline noise level. This simplified definition enables a consistent treatment of both single-component and multi-component structures, while remaining compatible with the more detailed hierarchical definitions presented in the literature.

Applying the above criteria to our dataset yields 315 burst-clusters containing a total of 360 bursts. The waiting time distribution of \frb \ is shown in\rev{ ~\AppFig{fig:waitingtime}.} Among these burst-clusters, 12.06\% display clear multi-component temporal profiles. Furthermore, 38 burst-clusters exhibit waiting times shorter than 123.8 ms, of which 34 consist of two peaks, three contain three peaks, and one features four adjacent peaks. To illustrate the fraction of multi-component structures among the total detected bursts, we imposed a threshold of at least five bursts per observing session; sessions with fewer than five bursts were excluded due to the relatively low burst rate of \frb. After this filtering, we obtained Figure~\ref{fig:Burst_number}. The bin size was chosen to be 50 days so as to ensure that each bin typically contains data from more than two observing sessions. Different colors represent bursts with different numbers of components. As shown, the fraction of multi-component burst-clusters remains at or below the 20\% level when calculated with a bin size of 50 days.

\begin{figure}[htbp]
    \centering
    \includegraphics[width=0.5\textwidth]{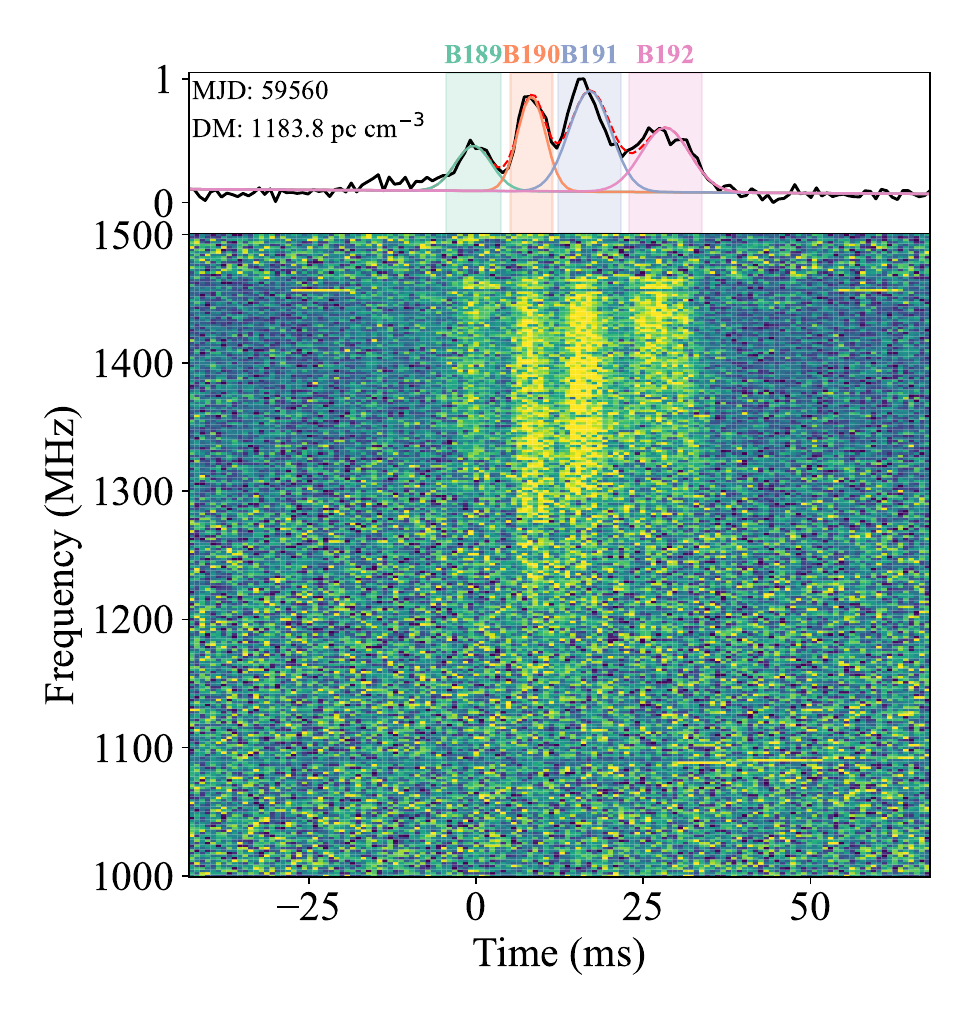}
    \caption{\textbf{Two-dimensional dynamic spectrum of a 4-component burst.} 
    \textbf{Top panel}: The average pulse profile, with the burst MJD and the measured DM value both annotated in the upper left corner. 
    Multiple Gaussian functions were fitted to the profile using curves of different colors, and the corresponding components are marked with the same colors. 
    \textbf{Bottom panel}: The waterfall plot of signal intensity (dynamic spectrum). }
    \label{fig:burst103}
\end{figure}

\section{Analysis}
\label{Analysis}
In this section, we analyze the multi-component structures of \frb \ through various aspects, including fluence and spectral analysis, as well as temporal analysis. We first examine the fluence distribution and spectral properties of the detected bursts, followed by a detailed temporal analysis of multi-component burst-clusters. Using the \rev{Lomb-Scargle (LS)} period search method, we identify potential periodicities in the burst profiles. Additionally, we apply linear fitting techniques to determine the time of arrival (TOA) of each subcomponent and assess the confidence of detected periodicities. 

\subsection{Fluence and Spectral Properties}
Figure~\ref{fig:fluence_distribution} shows the fluence distribution of the 316 detected bursts, after excluding 44 bursts that could not be energy-calibrated because no known-temperature noise signal was injected during those observations. From the figure, the fluence of multi-component burst-clusters does not appear to be more concentrated in any particular region, but rather tends to be consistent with the overall fluence distribution of all bursts.

\frb \ is a high-frequency, narrow-band source, with most bursts occurring near $1.4$ GHz and exhibiting relatively narrow bandwidths in FAST observations\citep{2024zhuyh}. As shown in Figure~\ref{fig:freq_bw_distribution}, the majority of bursts cluster around a center frequency of $1429.3 \pm 6.9$ MHz and a bandwidth of $112.1 \pm 8.6$ MHz. Multi-component burst-clusters, indicated by red pentagram markers, occupy a similar region in the frequency–bandwidth plane, suggesting that their spectral characteristics are broadly consistent with those of the full burst population. Statistical analysis, including the Kolmogorov–Smirnov (K-S) test and Mann–Whitney U test, confirms that neither the fluence distributions (p \textgreater \ 0.05 in both tests)~\citep{kolmogorov1933sulla, smirnov1948table, mann1947test} nor the center frequency and bandwidth distributions (K-S: p = 0.521; Mann–Whitney U: p = 0.685) of multi-component burst-clusters differ significantly from those of the overall burst sample, supporting the conclusion that their fluence and spectral properties are consistent with the general burst population.

\begin{figure*}[ht]
    \centering
    \includegraphics[width=0.8\linewidth]{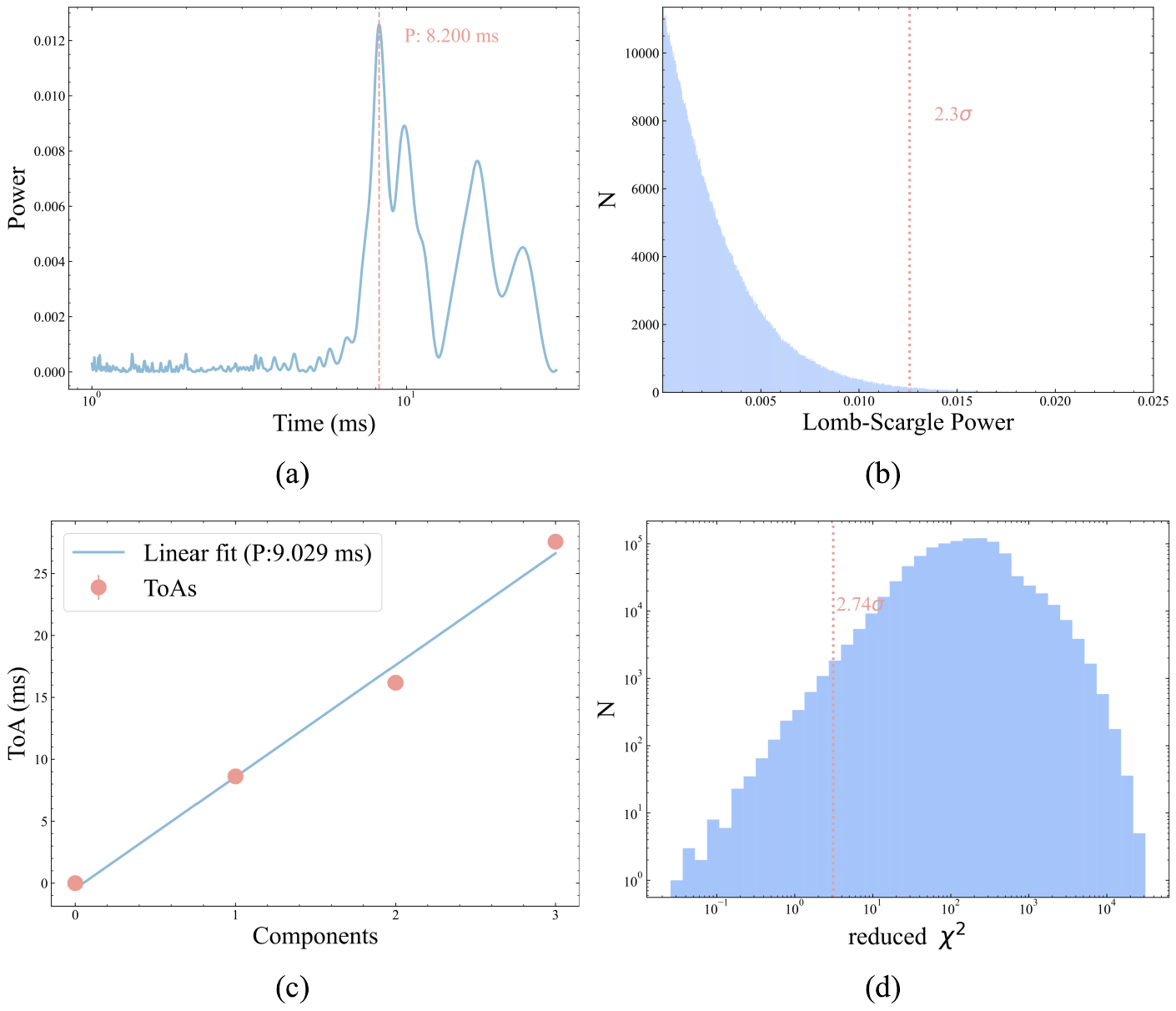}
    \caption{
    \textbf{Temporal analysis of the 4-component burst. }
    \textbf{(a)} Lomb–Scargle periodogram of the four-component burst. The horizontal axis represents the trial periods T (in milliseconds) scanned during the search, while the vertical axis shows the Lomb–Scargle power. The red dashed line indicates the period at which the power reaches its maximum, corresponding to the most significant periodicity.
    \textbf{(b)} Distribution of \rev{LS} powers obtained from $10^{6}$ randomly generated noise time series. 
    The horizontal axis shows the \rev{LS} power, and the vertical axis gives the occurrence count $N$. 
    The red dashed line marks the \rev{LS} power measured from the real burst data, corresponding to a significance level of $2.3\sigma$.
    \textbf{(c)} Linear fit of times of arrival (TOAs). 
    The horizontal axis represents the component number in the multi-component structure, 
    while the vertical axis shows the TOAs (ms). 
    The red points are the TOAs derived from Gaussian fits to each peak, 
    and the blue line is the linear fit, yielding the corresponding periodicity.
    \textbf{(d)} Significance test of the linear fit. 
    The figure shows the distribution of reduced $\chi^{2}$ values obtained from $10^{6}$ random trials, 
    where random time stamps were generated and compared with the previously derived linear fit to compute the reduced $\chi^{2}$. 
    The red dashed line marks the reduced $\chi^{2}$ value from the real data, 
    corresponding to a confidence level of $2.74\sigma$.
    }
    \label{fig:Temporal analysis}
\end{figure*}

\subsection{Temporal Analysis of Multi-component Burst-clusters}
In previous studies, periodicity searches of multi-component structures have been considered highly important. 
In \citet{2022NJR}, the periodicities of multi-component burst-clusters from FRB~20201124A were systematically analyzed, and a distribution of characteristic periods was obtained.
\citet{2024kramer} further suggested that such periodicities might be related to the rotational period of the source. 
Motivated by these results, we performed a periodicity search for the multi-component structures of \frb. 
Since meaningful periodicity searches can only be conducted for bursts with at least three components, 
we applied two methods to one 4-component burst and three 3-component bursts in our dataset. 
Figure~\ref{fig:burst103} presents the 4-component burst-cluster. 
The top panel shows the burst profile, where we fitted multiple Gaussian functions to identify each component, highlighted with different colors. 
The bottom panel displays the corresponding waterfall plot of the same burst-cluster. 
The periodicity analysis for this 4-component burst-cluster is shown in Figure~\ref{fig:Temporal analysis}, 
while the results for the 3-component bursts are provided in \rev{the ~\AppFig{fig:3-component burst} and \AppFig{fig:burst_series}).}

\subsubsection{Lomb–Scargle Periodicity Search}
Compared to traditional Fourier transform techniques, \rev{the LS method} is more effective for irregularly sampled data and is particularly useful in extracting significant periodic signals from pulse profiles.\rev{The LS method} has become a widely adopted tool for period analysis.

We processed the data of the 4-component burst-cluster as follows. 
First, the signals across all frequency channels were averaged 
(after correcting for the dispersion measure and removing RFI contamination) 
to obtain the time-domain pulse profile. 
This profile was then normalized to yield a one-dimensional time series. 
We applied the \rev{LS} algorithm to this series, 
where the searched periods depend on both the length of the time series and the number of trial periods. 
For this analysis, we restricted the search range to 1--30\,ms. 
As shown in panel (a) of Figure~\ref{fig:Temporal analysis}, the \rev{LS} power spectrum of a representative burst with four distinct subcomponents exhibits a prominent peak at \( P = 8.200^{+0.686}_{-0.568}\ \mathrm{ms} \), with additional peaks corresponding to harmonic multiples of the fundamental period. The quoted asymmetric uncertainties are derived from the full width at half maximum (FWHM) of the main peak.

To assess the statistical significance of this result, we generated \(10^6\) synthetic noise profiles based on the original pulse time series. For each realization, we computed the \rev{LS} power at the period where the original profile exhibited its highest \rev{LS} power. The significance was evaluated by comparing the distribution of these \rev{LS} powers with the observed peak power in the original data. The result, corresponding to a confidence level of about 2.3 \(\sigma\), is shown in panel (b) of Figure~\ref{fig:Temporal analysis}.

\subsubsection{Linear Fitting of Sub-pulse Arrival Times (TOA) for Period Determination}
As previously outlined, \cite{2022andersen} introduced a method to determine periodicity by extracting the time of arrival (TOA) of each subcomponent via multi-Gaussian fitting. These TOAs are then fitted linearly using the following model:
\begin{equation}
    t_i = \bar{d}n_i + T_0,
\end{equation}
where \( t_i \) represents the TOA of the \(i\)-th subcomponent, \( \bar{d} \) denotes the average spacing between subcomponents within a single burst (which corresponds to the target period \(P\)), \( n_i \) is the index of the \(i\)-th subcomponent, and \( T_0 \) is the TOA of the first component. The result of this linear fit, shown in panel (c) of Figure~\ref{fig:Temporal analysis}, yields a periodicity of \(P = 9.03 \pm 0.56\ \mathrm{ms}\).

\begin{revblock}
    
To evaluate the robustness of the detected periodicity, we generated the subcomponent spacings $d_i$ from a shifted–exponential law (i.e., a Poisson-process inter-arrival distribution with a deterministic offset),
\begin{equation}
d_i = \bar d\,\big[\eta - (1-\eta)\ln(1-u_i)\big],\qquad
u_i \sim \mathcal{U}(0,1),
\end{equation}
with shift fraction $\eta = 0.2$ following \citet{Pasto_2023}.
For each simulated sample, we computed the reduced chi-squared ($\chi^2_{\rm red}$) and assessed significance by comparing the observed $\chi^2_{\rm red}$ against the simulation distribution.
The observed periodicity corresponds to a confidence level of $2.74\sigma$, as shown in panel~(d) of Figure~\ref{fig:Temporal analysis}.
\end{revblock}

\begin{table*}[ht]
    \centering
    \caption{Statistics of multi-component burst-clusters from selected FRB sources}
    \begin{tabular}{l|cccc}
        \toprule
        \textbf{FRB Source} & \textbf{Total burst-clusters} & \textbf{Multi-component Burst-clusters ($\geq$2)} & \textbf{Fraction} & \textbf{Burst Rate (h$^{-1}$)} \\ 
        \midrule
        \frb & 315  & 38   & 12.06\% & 5 \\
        FRB 20201124A & 604  & 186  & 30.79\% & 46.5 \\
        FRB 20121102A & 1312 & 292  & 22.26\% & 27.7 \\
        FRB 20240114A & 8778 & 1858 & 21.17\% & 333 \\
        \bottomrule
    \end{tabular}
    \label{table:statistics of multi-component}
\end{table*}

\section{Results}
\label{Result}
\label{sect:Result}

In this study, we analyzed the fluence, spectral, and temporal properties of multi-component burst-clusters in the \frb\ dataset. We find that the fluence and spectral distributions of the bursts in multi-component burst-clusters are statistically consistent with the overall burst population, while temporal analyses using \rev{LS} periodograms and linear fitting of sub-pulse TOAs reveal candidate periodicities in bursts with three or more components, with their significance evaluated through extensive noise simulations. The temporal analyses of the other three-component burst-clusters also yield corresponding quasi-periodicities. However, no clear correlation is found among these periods, which is consistent with the results of \citet{2022NJR} for the multi-component burst-clusters of FRB~20201124A.

We attempted to connect the periodicities obtained from the multi-component burst-clusters with the intrinsic rotation period of the source \citep{2024kramer}. Using the relation $P_{\mu} = (0.94 \pm 0.04) \times P ^{(+0.97 \pm 0.05)}\,\mathrm{ms}$, we derived the corresponding rotation period and performed phase-folding of the time series. However, this approach did not yield satisfactory results. Furthermore, when we directly folded the TOAs sequence of all 360 bursts using the periodicities inferred from the multi-component burst-clusters, no evidence of a significant periodic signal was found.

To quantify and compare the multiplicity of emission within clusters across active repeating FRBs, we focus on four well-studied sources: FRB~20121102A \citep{2021Lidi}, FRB~20190520B (this work), FRB~20240114A \citep{2025ZhangJS}, and FRB~20201124A \citep{2022zhangyk}. For consistency, we adopt unified working definitions: a \textit{burst-cluster} is defined by a maximum intra-cluster time separation set by the valley between the two peaks of the bimodal waiting-time distribution (following previous studies), while an individual \textit{burst} within a cluster is identified as a component whose peak-to-valley flux contrast exceeds $3\sigma$ above the baseline noise level, consistent with Section~\ref{DATA}. The waiting-time distribution used for this definition is shown in Figure~\ref{fig:waitingtime}.

\begin{figure}[b]
    \centering
    \includegraphics[width=1\linewidth]{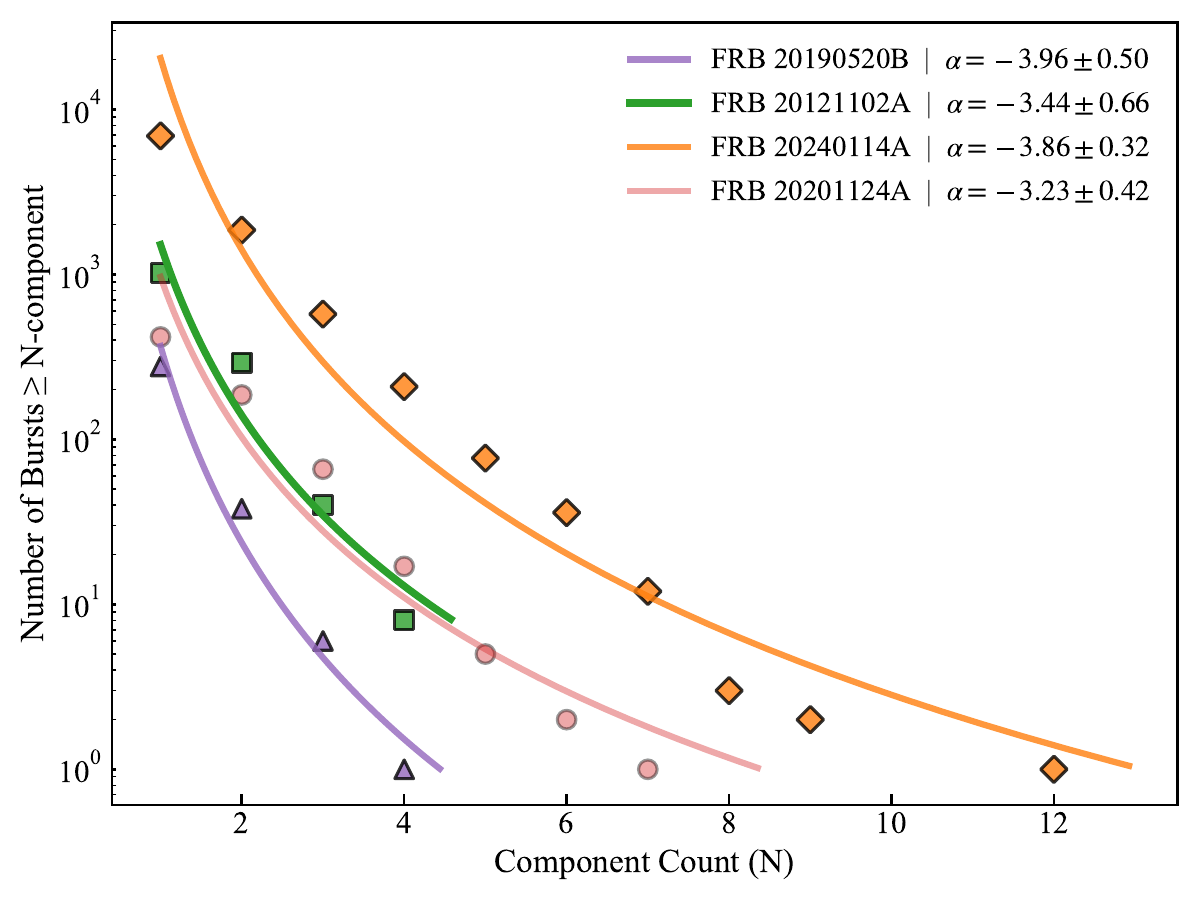}
    \caption{Cumulative distribution of burst-clusters as a function of the number of components ($n$) per cluster for \frb\ and other repeating FRB sources. The vertical axis shows the number of clusters with $n$ or more components, and the horizontal axis represents $n$.}
    \label{fig:cumulative_distri}
\end{figure}

\begin{figure}[t]
    \centering
    \includegraphics[width=1\linewidth]{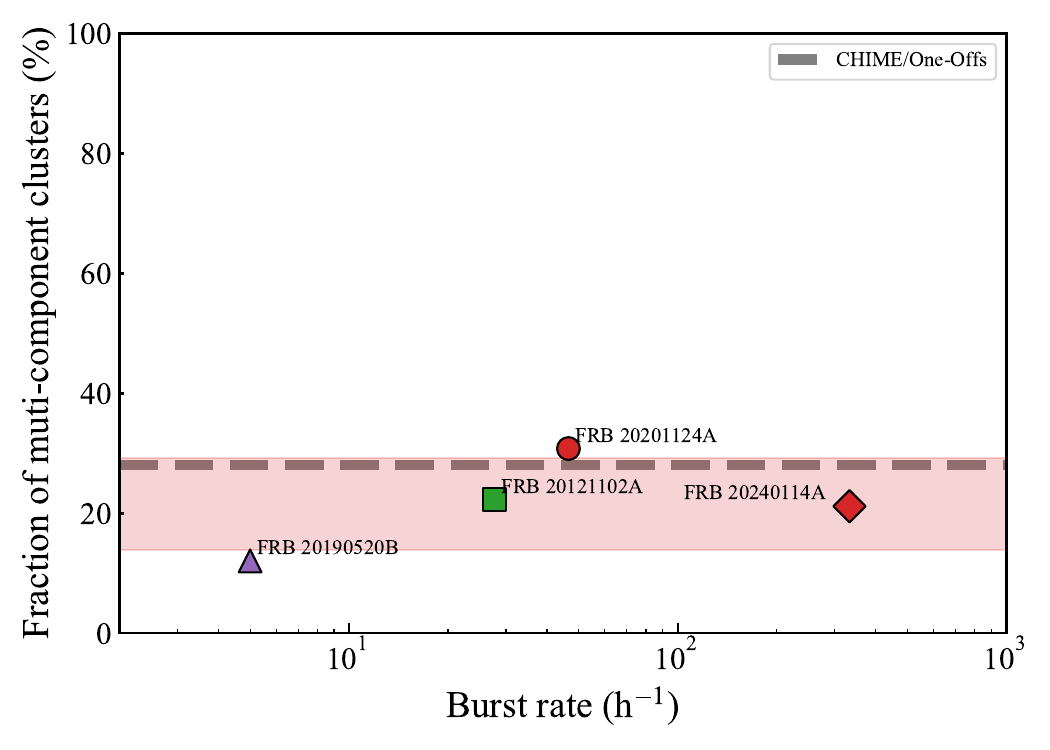}
    \caption{\textbf{Relation between the burst rate and the fraction of multi-component burst-clusters for different FRB sources.} Different colors and marker shapes denote different FRB sources. The red-shaded area marks the range of the fraction. The gray dashed line represents the CHIME result from visual re-examination reported by \citet{2025Pastor}, corresponding to a fraction of about 28\%.}
    \label{fig:rate&Fraction}
\end{figure}

Table~\ref{table:statistics of multi-component} summarizes the statistics of multi-component burst-clusters for the four repeating FRBs analyzed in this work. 
In addition, Figure~\ref{fig:rate&Fraction} presents the fractions of multi-component structures and the corresponding burst rates for each source. 
FRB~20190520B shows a total of 315 burst-clusters, of which only 38 exhibit multiple components, corresponding to a fraction of 12.06\% and the lowest burst rate among the four sources. 
In contrast, FRB~20201124A, FRB~20121102A, and FRB~20240114A display substantially higher fractions of multi-component burst-clusters, at 30.79\%, 22.26\%, and 21.17\%, respectively. 
Notably, FRB~20201124A has the highest fraction of multi-component burst-clusters, but without the highest burst rate. 
This indicates that the burst rate and the fraction of multi-component structures are not necessarily correlated.

\citet{2025Pastor} reported that 7 out of 24 one-off FRBs in their Apertif sample ($\sim$30\%) exhibit multi-component temporal structures. At the same time, they also re-examined 130 one-off FRBs from the First CHIME/FRB Catalog and found that about 28\% display complex, multi-component morphologies. These comparable fractions suggest that the occurrence of multi components is not uncommon even among apparently one-off FRBs. The authors further argued that such complexity may reflect an intrinsic property of FRB emission, in close analogy with pulsars, whose profiles frequently show multi-component structures as a consequence of their underlying radiation mechanisms.

For each source, we construct the cumulative distribution of component counts per cluster, plotting $N_{\mathrm{bursts}}(\geq N)$ versus $N$ in semi–log space and fitting power-law models $N_{\mathrm{bursts}}(\geq N)\propto N^{-\alpha}$ (slopes listed in the legend of Figure~\ref{fig:cumulative_distri}). The fitting was performed by ordinary least-squares regression in semi–log space, with best-fit parameters and 1\(\sigma\) uncertainties derived from the covariance matrix. The power-law distribution indicates a scale-free behavior in the burst emission. Such a scale-free distribution suggests that there is no characteristic number of sub-pulses, with both small and large sub-pulse clusters arising from the same underlying process. This behavior is naturally consistent with a self-organized criticality (SOC) scenario, where the FRB source evolves toward a critical state rather than being governed by purely random or Poissonian processes.

\frb \ and FRB~20121102A exhibit steeper slopes, indicating that clusters with large $N$ are comparatively rare, whereas FRB~20240114A and FRB~20201124A show shallower slopes, implying a higher fraction of clusters with multiple components. These contrasting behaviors may reflect intrinsic differences in emission mechanisms or source environments among repeaters. In principle, a single trigger event can launch multiple particle beams propagating along different magnetic field lines; however, the geometry dictated by the magnetospheric configuration and the star's rotation prevents not all resulting FRB beams from crossing the observer's line of sight, which causes the sub-pulses to exhibit identical statistical properties~\citep{2019WWYApjL}.

\section{Discussion and Conclusion}
\label{sec:Dissusion}
In this work, we carried out a systematic investigation of the multi-component burst-clusters from FRB~20190520B using high-sensitivity FAST observations. Our fluence and spectral analyses show that multi-component clusters share broadly similar properties with single-component bursts, with no significant differences in flux density distribution, center frequency, or bandwidth. These results indicate that the occurrence of multi-component structures is not driven by distinct emission energetics or spectral characteristics, but instead arises from temporal modulation within the same underlying emission process.

The temporal analysis further reveals that several bursts exhibit quasi-periodic sub-structures on millisecond timescales. By applying both Lomb–Scargle periodograms and linear fits to sub-pulse times of arrival, we identified multiple distinct quasi-periodicities within the multi-component structures. Multi-component structures are observed to occur sporadically and exhibit variations from burst to burst, in agreement with previous studies of other FRBs. \rev{This behavior suggests that} such fine-scale structures are more likely linked to intrinsic magnetospheric emission processes. However, no corresponding spin period can be established across the overall sample.

Building on these results, we carried out a broader comparison across different repeating FRBs. Our comparative analysis of multi-component burst-clusters highlights several key results. First, FRB~20190520B exhibits a markedly lower fraction of multi-component clusters ($\sim$12\%) compared to other active repeaters such as FRB~20201124A, FRB~20121102A, and FRB~20240114A. Second, we find no evidence for a correlation between the overall burst rate and the fraction of multi-component structures, suggesting that the emergence of such complexity is governed primarily by intrinsic emission processes rather than the global activity level. 

Third, comparisons with one-off FRBs \footnote{\rev{There is ongoing debate on whether all FRBs are capable of repetition with widely varying activity rates~\citep{2019Ravi,2024Kirsten,2024MNYamasaki}.}}reveal that $\sim$30\% also display multi-component structures, implying that such morphologies are not unique to repeaters but may reflect a more fundamental property of FRB emission, analogous to the multi-component profiles seen in pulsars.
Finally, the cumulative distribution of component counts follows a power-law form, pointing toward a scale-free behavior such as \rev{SOC}. This finding challenges purely random interpretations and links FRBs to a broader class of astrophysical SOC phenomena, such as solar flares and \rev{starquakes}, thereby offering new physical perspectives on FRB mechanisms.

Future progress will benefit from larger samples of multi-component burst-clusters across different repeaters, as well as higher time-resolution observations that can better resolve microsecond-level structures. In particular, polarimetric and wideband studies may reveal whether the temporal regularity observed in some bursts is accompanied by correlated changes in polarization or frequency drift. Combining such analyses with theoretical modeling of magnetospheric plasma processes will be crucial to uncover the physical origin of FRB multi-component structure.

\section*{Acknowledgments}
This work was supported by the National Natural Science Foundation of China (NSFC) under Programs No. 11988101 and 12203069. C.H.N. acknowledges support from the National SKA Program of China (2022SKA0130100), the Basic Research Project of Central China Normal University, and the Hubei QB Project. C.H.N. also thanks the CAS Youth Interdisciplinary Team and the Foundation of Guizhou Provincial Education Department for Grant No. KY(2023)059.

This work made use of data from FAST (Five-hundred-meter Aperture Spherical Radio Telescope). We thank the FAST data center operations team for scheduling the follow-up observations and facilitating data acquisition.

\bibliography{main}{}
\bibliographystyle{aasjournal}

\appendix
\restartappendixnumbering       
\renewcommand{\figurename}{Appendix Figure}
\renewcommand{\tablename}{Appendix Table}

\setcounter{figure}{0}
\setcounter{table}{0}

\renewcommand{\thefigure}{A\arabic{figure}}
\renewcommand{\thetable}{A\arabic{table}}
\begin{longtable}{c|cccccc}
\caption{\textbf{Properties of the multi-component burst-clusters ($\ge$2-components).}} \\
\hline
Burst ID & Component & TOAs & DM & Center Freq & Bottom Freq & Top Freq \\
& & (MJD)@1.5 GHz & (pc\,cm$^{-3}$) & (MHz) & (MHz) & (MHz) \\
\hline
\endfirsthead
\hline
\endhead
        B88 & 1 & 59172.20155284418433 & 1201.3(0.1) & 1439.596752 & 1411.179314 & 1468.014189 \\ 
        B89 & 2 & 59172.20155299289763 & 1201.3(0.1) & 1444.22042 & 1374.40458 & 1499.87793 \\ 
        B91 & 1 & 59208.08742101160169 & 1195.7(0.1) & 1445.429289 & 1382.781953 & 1499.87793 \\ 
        B92 & 2 & 59208.08742137546506 & 1195.7(0.1) & 1429.833983 & 1342.71176 & 1499.87793 \\ 
        B104 & 1 & 59208.11264542803110 & 1193.2(0.2) & 1440.03506 & 1385.75797 & 1494.31214 \\ 
        B105 & 2 & 59208.11264683925401 & 1193.2(0.2) & 1415.413442 & 1388.229317 & 1442.597568 \\ 
        B116 & 1 & 59224.04922651168454 & 1198.2(0.1) & 1453.823954 & 1340.785357 & 1499.87793 \\ 
        B117 & 2 & 59224.04922798402549 & 1198.2(0.1) & 1418.567731 & 1332.101247 & 1499.87793 \\ 
        B134 & 1 & 59306.82174683937774 & 1194.1(0.1) & 1393.457246 & 1304.941428 & 1481.973063 \\ 
        B135 & 2 & 59306.82174697117443 & 1194.1(0.1) & 1392.749582 & 1316.966929 & 1468.532235 \\ 
        B140 & 1 & 59306.83829678239999 & 1194.0(0.1) & 1413.714683 & 1364.411655 & 1463.017712 \\ 
        B141 & 2 & 59306.83829764356051 & 1194.0(0.1) & 1371.266031 & 1232.997796 & 1499.87793 \\ 
        B152 & 1 & 59347.73112721699727 & 1205.2(0.7) & 1448.46554 & 1408.07488 & 1488.8562 \\ 
        B153 & 2 & 59347.73112785528065 & 1205.2(0.7) & 1392.14125 & 1326.61413 & 1457.66838 \\ 
        B154 & 1 & 59355.70649803764536 & 1196.9(0.2) & 1379.74938 & 1320.75878 & 1438.739981 \\ 
        B155 & 2 & 59355.70649877208052 & 1196.9(0.2) & 1359.894877 & 1284.168482 & 1435.621271 \\ 
        B159 & 1 & 59475.37105939972389 & 1190.4(0.1) & 1408.092101 & 1307.478778 & 1499.87793 \\ 
        B160 & 2 & 59475.37105984813388 & 1190.4(0.1) & 1339.230423 & 1287.136193 & 1391.324653 \\ 
        B161 & 1 & 59475.38309604497044 & 1191.4(0.2) & 1447.197391 & 1376.314962 & 1499.87793 \\ 
        B162 & 2 & 59475.38309652624594 & 1191.4(0.2) & 1447.197391 & 1376.314962 & 1499.87793 \\ 
        B178 & 1 & 59554.15544217405113 & 1184.8(0.1) & 1441.920893 & 1361.395049 & 1499.87793 \\ 
        B179 & 2 & 59554.15544230146770 & 1184.8(0.1) & 1360.228531 & 1228.823934 & 1491.633128 \\ 
        B189 & 1 & 59560.16473174953717 & 1183.8(0.1) & 1488.30718 & 1404.33824 & 1499.87793 \\ 
        B190 & 2 & 59560.16473193652928 & 1183.8(0.1) & 1386.329255 & 1238.535024 & 1499.87793 \\ 
        B191 & 3 & 59560.16473212611891 & 1183.8(0.1) & 1382.744672 & 1278.593328 & 1486.896016 \\ 
        B192 & 4 & 59560.16473236353340 & 1183.8(0.1) & 1398.097429 & 1303.922838 & 1492.27202 \\ 
        B209 & 1 & 59580.08211609908903 & 1180.7(0.1) & 1475.632947 & 1417.787545 & 1499.87793 \\ 
        B210 & 2 & 59580.08211673732876 & 1180.7(0.1) & 1368.759931 & 1252.668636 & 1484.851225 \\ 
        B211 & 1 & 59580.08666774126323 & 1182.8(0.2) & 1357.036045 & 1282.381258 & 1431.690833 \\ 
        B212 & 2 & 59580.08666789661947 & 1182.8(0.2) & 1417.081412 & 1376.023966 & 1458.138859 \\ 
        B219 & 1 & 59604.00552298274852 & 1185.0(0.1) & 1215.353104 & 1133.863568 & 1296.842639 \\ 
        B220 & 2 & 59604.00552331320068 & 1185.0(0.1) & 1242.040473 & 1143.58 & 1340.500946 \\ 
        B221 & 3 & 59604.00552347524354 & 1185.0(0.1) & 1301.10519 & 1171.495354 & 1430.715025 \\ 
        B229 & 1 & 59632.93453996753669 & 1194.3(0.1) & 1420.368541 & 1325.873149 & 1499.87793 \\ 
        B230 & 2 & 59632.93454041893710 & 1194.3(0.1) & 1447.96008 & 1426.09629 & 1469.82387 \\ 
        B231 & 3 & 59632.93454092050524 & 1194.3(0.1) & 1219.488262 & 1160.218376 & 1278.758148 \\ 
        B233 & 1 & 59641.92190177011798 & 1185.0(0.0) & 1421.9542 & 1361.72358 & 1482.18482 \\ 
        B234 & 2 & 59641.92190313873289 & 1185.0(0.0) & 1421.13769 & 1323.59275 & 1499.87793 \\ 
        B252 & 1 & 59671.82928816841741 & 1180.6(0.2) & 1245.522078 & 1136.090558 & 1354.953599 \\ 
        B253 & 2 & 59671.82928860051470 & 1180.6(0.2) & 1134.2553 & 1087.83086 & 1180.67974 \\ 
        B258 & 1 & 59671.85351685265050 & 1178.0(0.1) & 1436.581926 & 1342.090273 & 1499.87793 \\ 
        B259 & 2 & 59671.85351746199740 & 1178.0(0.1) & 1448.52009 & 1405.302996 & 1491.737183 \\ 
        B261 & 1 & 59671.85659109426342 & 1176.2(0.2) & 1263.827493 & 1184.583878 & 1343.071107 \\ 
        B262 & 2 & 59671.86551290835632 & 1176.2(0.2) & 1246.42978 & 1169.119366 & 1323.740194 \\ 
        B263 & 1 & 59671.86299286691064 & 1181.1(0.1) & 1350.610982 & 1287.639805 & 1413.582159 \\ 
        B264 & 2 & 59671.86299317443627 & 1181.1(0.1) & 1357.75772 & 1295.977996 & 1419.537445 \\ 
        B265 & 3 & 59671.86299358931137 & 1181.1(0.1) & 1209.754147 & 1000 & 1438.069033 \\ 
        B267 & 1 & 59671.86551252580102 & 1184.3(0.2) & 1324.79933 & 1238.78023 & 1410.98673 \\ 
        B268 & 2 & 59672.82162277348107 & 1184.3(0.2) & 1091.97256 & 1000 & 1186.70003 \\ 
        B284 & 1 & 59673.85253208376525 & 1186.5(0.1) & 1261.634547 & 1193.052185 & 1330.216909 \\ 
        B285 & 2 & 59673.85253500939143 & 1186.5(0.1) & 1345.162549 & 1235.393067 & 1454.932032 \\ 
        B301 & 1 & 59705.75039521722647 & 1187.7(0.1) & 1370.16629 & 1347.01602 & 1393.31656 \\ 
        B302 & 2 & 59705.75039591296081 & 1187.7(0.1) & 1450.188561 & 1430.261616 & 1470.115507 \\ 
        B308 & 1 & 59747.64526278508856 & 1183.0(0.1) & 1385.97266 & 1323.75721 & 1448.18811 \\ 
        B309 & 2 & 59747.64526328403008 & 1183.0(0.1) & 1444.54027 & 1397.67329 & 1491.40724 \\ 
        B328 & 1 & 59820.43713358692185 & 1172.8(0.1) & 1386.205173 & 1306.050971 & 1466.359375 \\ 
        B329 & 2 & 59820.43713382320857 & 1172.8(0.1) & 1295.904172 & 1224.263213 & 1367.54513 \\ 
        B343 & 1 & 59839.38941793060076 & 1164.5(0.1) & 1428.233911 & 1369.898468 & 1486.569353 \\ 
        B344 & 2 & 59839.38941815464932 & 1164.5(0.1) & 1427.561852 & 1380.977244 & 1474.146461 \\ 
        B348 & 1 & 59852.33670939735021 & 1162.2(0.3) & 1422.930849 & 1362.013439 & 1483.84826 \\ 
        B349 & 2 & 59852.33671153239993 & 1162.2(0.3) & 1320.748105 & 1200.447395 & 1441.048814 \\ 
        B352 & 1 & 59871.30124151545169 & 1176.4(0.1) & 1322.34169 & 1300.34506 & 1344.33832 \\ 
        B353 & 2 & 59871.30124189160415 & 1176.4(0.1) & 1338.116295 & 1250.127555 & 1426.105036 \\ 
        B361 & 1 & 59894.21710628105211 & 1162.8(0.2) & 1417.477493 & 1364.383841 & 1470.571145 \\ 
        B362 & 2 & 59894.21710658693337 & 1162.8(0.2) & 1419.411757 & 1372.745718 & 1466.077797 \\ 
        B373 & 1 & 59910.16238363894081 & 1169.1(0.0) & 1277.417033 & 1118.077015 & 1436.75705 \\ 
        B374 & 2 & 59910.16239624022273 & 1169.1(0.0) & 1417.505606 & 1372.177468 & 1462.833743 \\ 
        B376 & 1 & 59910.16408148084884 & 1179.5(0.1) & 1416.42458 & 1341.516099 & 1491.333061 \\ 
        B377 & 2 & 59910.16408181364386 & 1179.5(0.1) & 1426.596673 & 1329.984854 & 1499.87793 \\ 
        B379 & 1 & 59910.16880359954666 & 1172.7(0.1) & 1452.655465 & 1383.047263 & 1499.87793 \\ 
        B380 & 2 & 59910.16880368578859 & 1172.7(0.1) & 1428.3993 & 1373.070936 & 1483.727665 \\ 
        B384 & 1 & 59910.17863870527799 & 1179.3(0.1) & 1465.37969 & 1403.50529 & 1500 \\ 
        B385 & 2 & 59910.17863917985960 & 1179.3(0.1) & 1305.5432 & 1226.72705 & 1384.35934 \\ 
        B410 & 1 & 59960.06469329555694 & 1177.9(0.1) & 1446.11881 & 1409.30689 & 1482.93073 \\ 
        B411 & 2 & 59960.06469358679897 & 1177.9(0.1) & 1448.65169 & 1406.49364 & 1490.80974 \\ 
        B413 & 1 & 59960.06738932633743 & 1181.8(0.1) & 1428.316426 & 1389.341124 & 1467.291727 \\ 
        B414 & 2 & 59960.06738956437766 & 1181.8(0.1) & 1346.340465 & 1248.203413 & 1444.477516 \\ 
        B421 & 1 & 59987.95267183618125 & 1171.9(0.1) & 1330.643188 & 1248.401412 & 1412.884963 \\ 
        B422 & 2 & 59987.95267201295064 & 1171.9(0.1) & 1275.814435 & 1203.131564 & 1348.497306 \\
\hline
\end{longtable}

\begin{figure*}[h]
    \centering
    \gridline{
        \fig{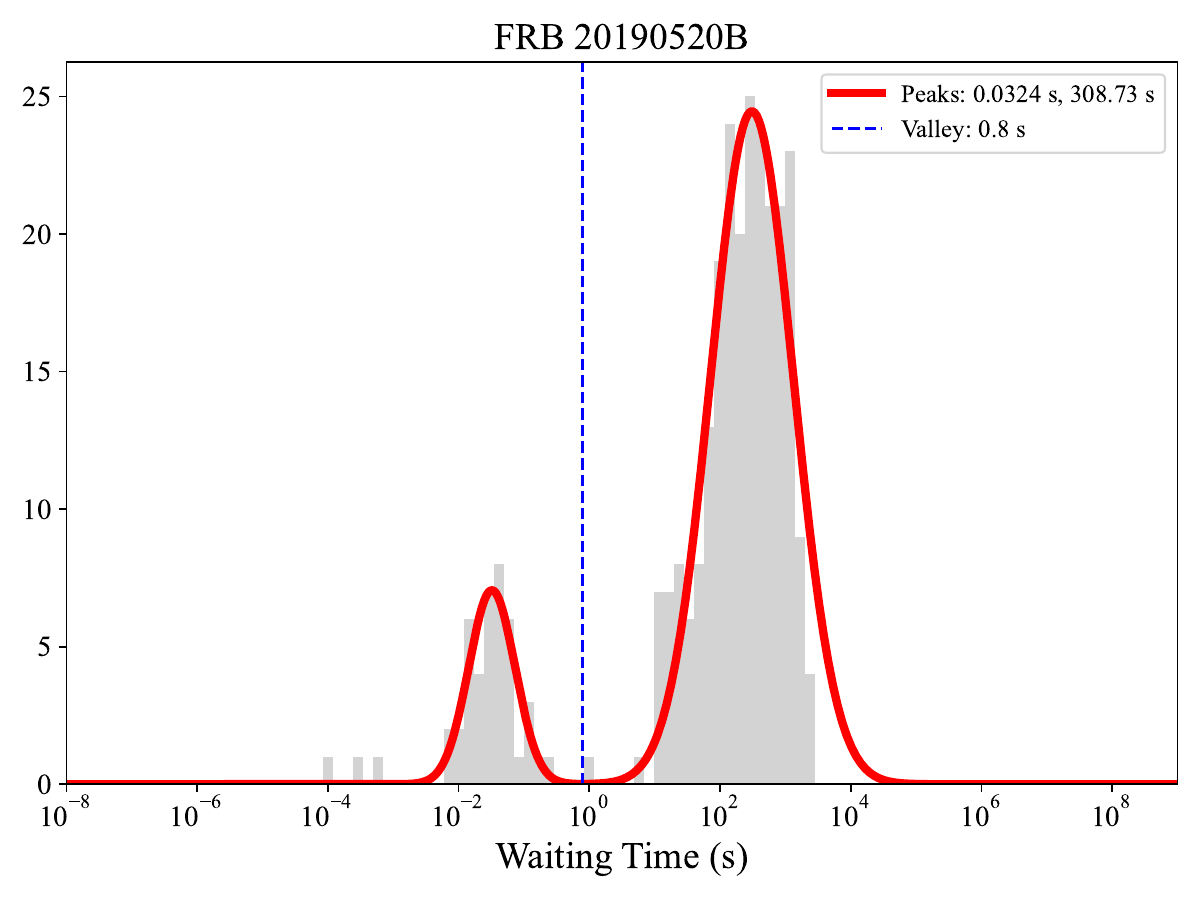}{0.45\textwidth}{a}
        \fig{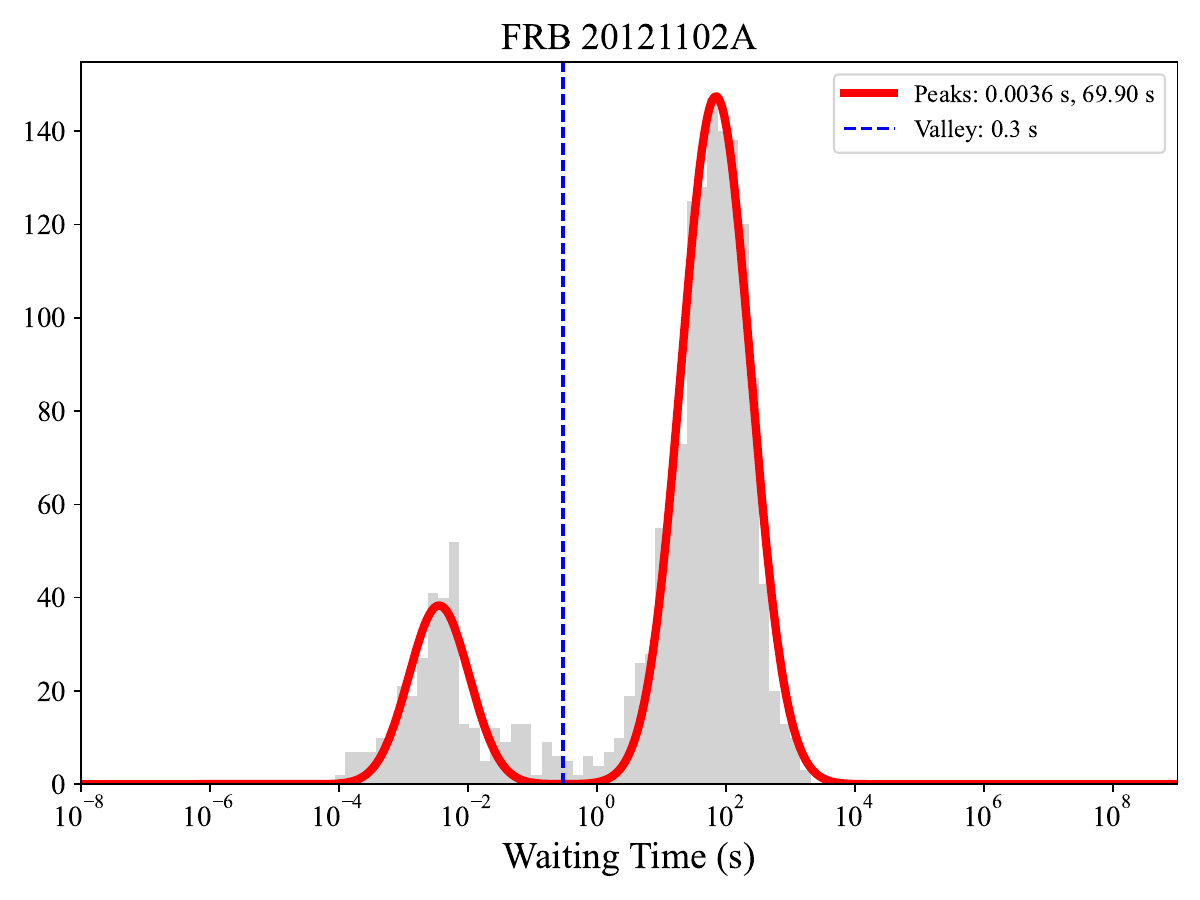}{0.45\textwidth}{b}
    }
    \gridline{
        \fig{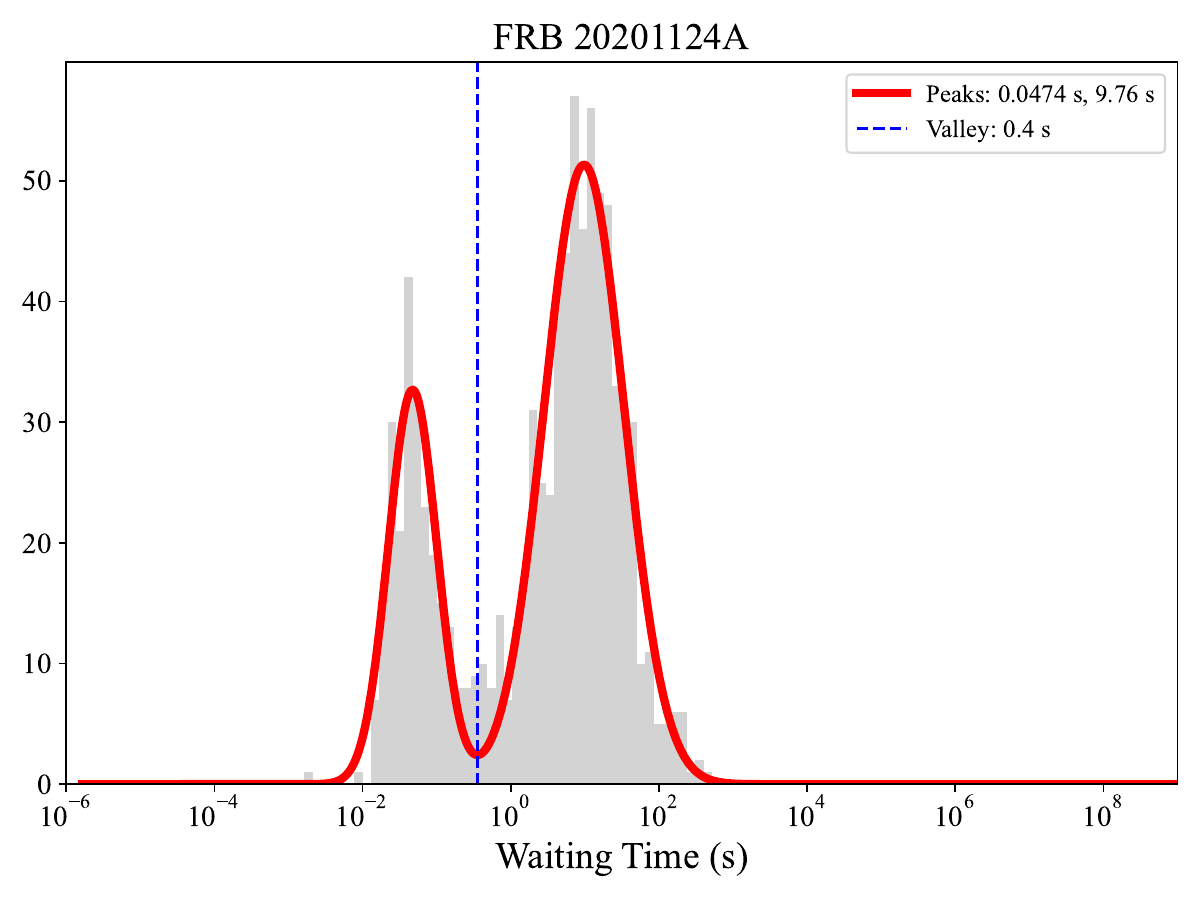}{0.45\textwidth}{c}
        \fig{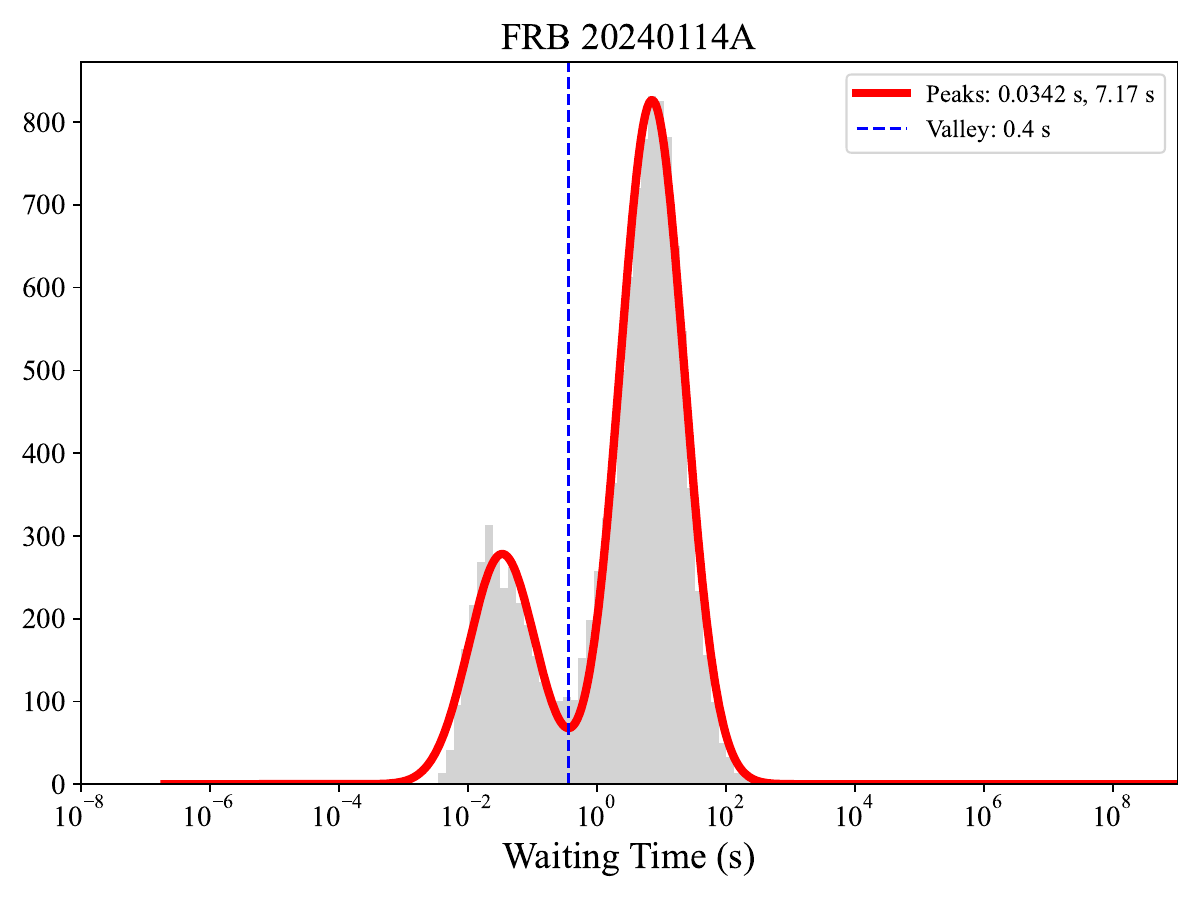}{0.45\textwidth}{d}
    }
    \caption{\textbf{Waiting time distribution of four FRBs.} The histogram shows the waiting time distribution, with the red line representing the lognormal double-peak fit. The two peaks correspond to the characteristic waiting times, with their values labeled in the upper right corner of the plot. The blue dashed line marks the valley between the peaks, and the valley time is used as a criterion to distinguish whether the events belong to a single burst-cluster.
    }
    \label{fig:waitingtime}
\end{figure*}

\begin{figure*}[htbp]
    \centering
    \gridline{
        \fig{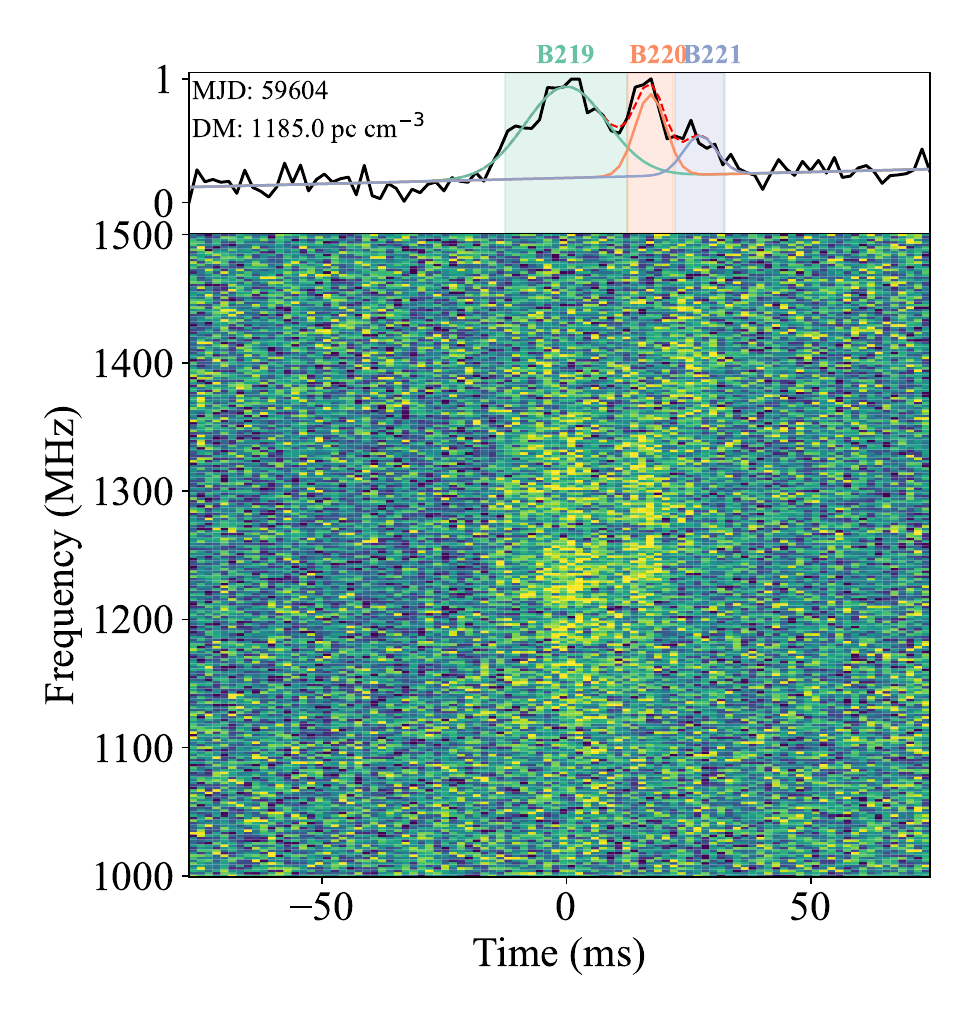}{0.29\textwidth}{\ \ \ \ 3-1}
        \fig{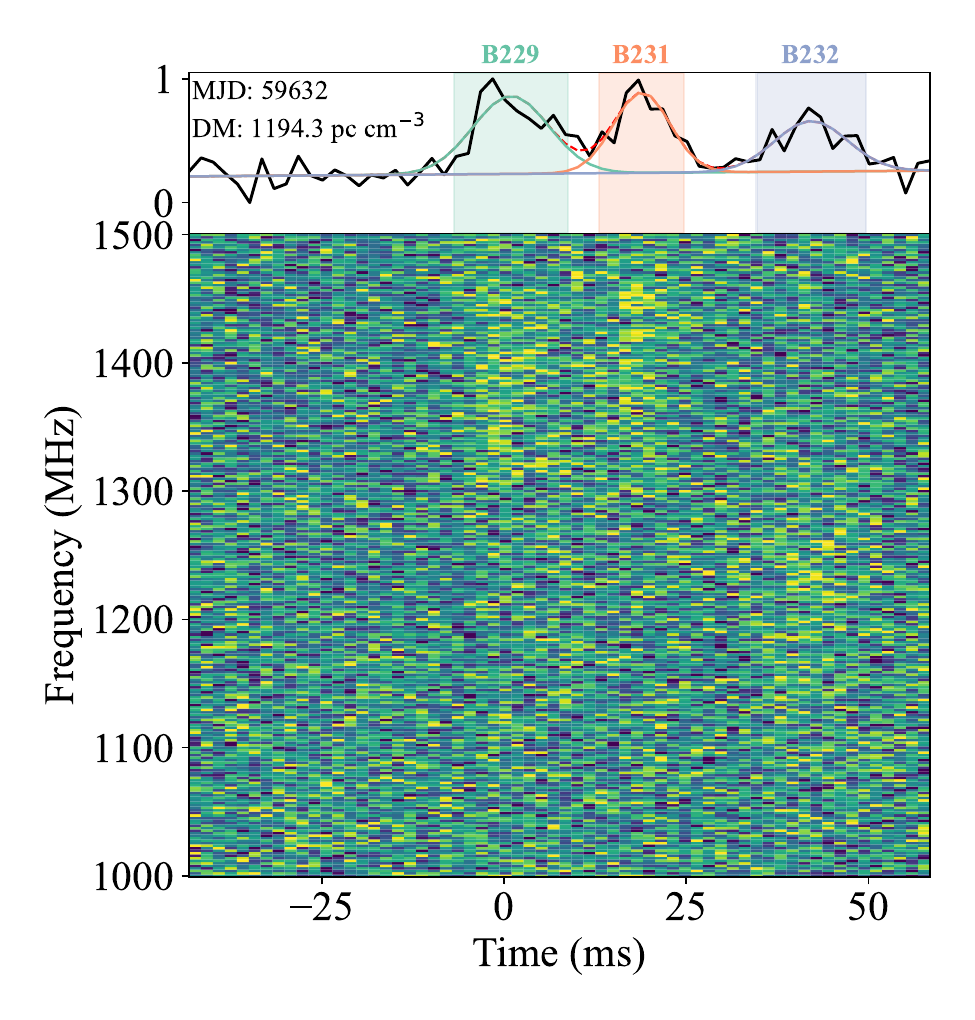}{0.29\textwidth}{\ \ \ \ 3-2}
        \fig{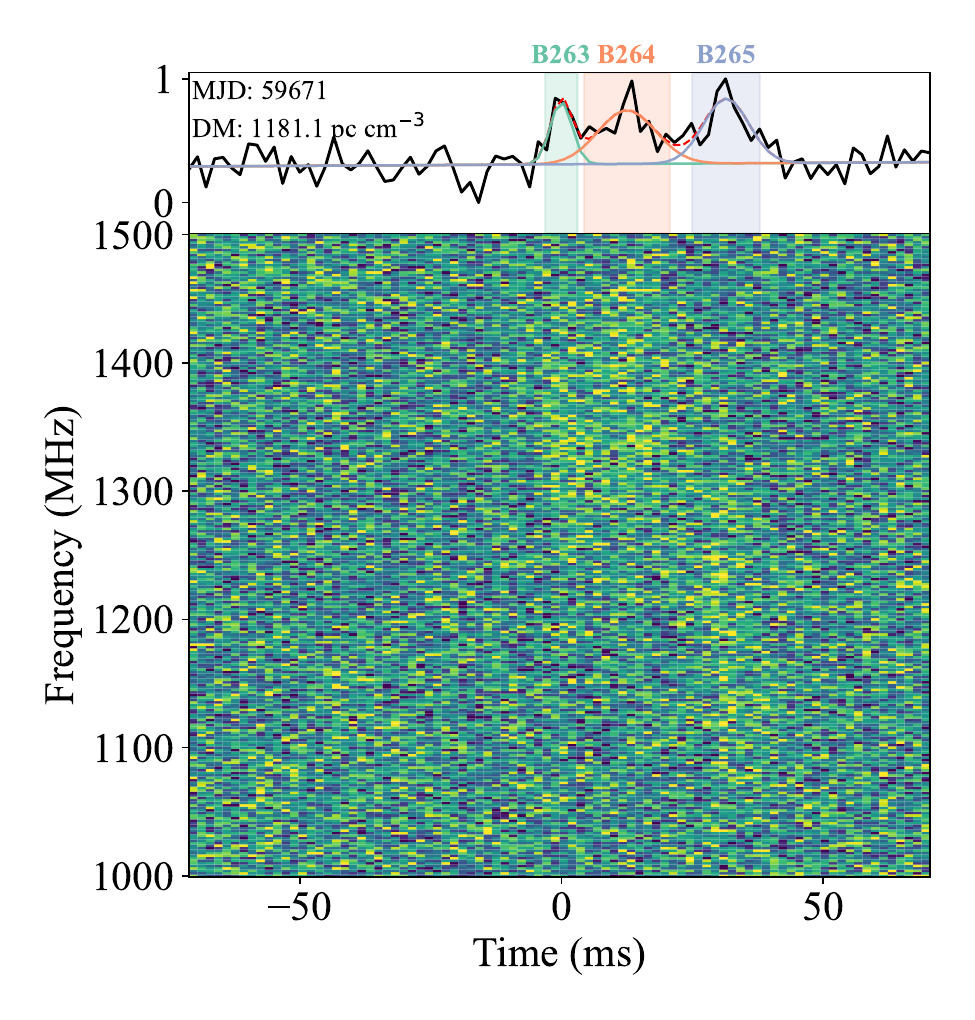}{0.29\textwidth}{\ \ \ \ 3-3}
    }
    \vspace{0 cm}   

    \gridline{
        \fig{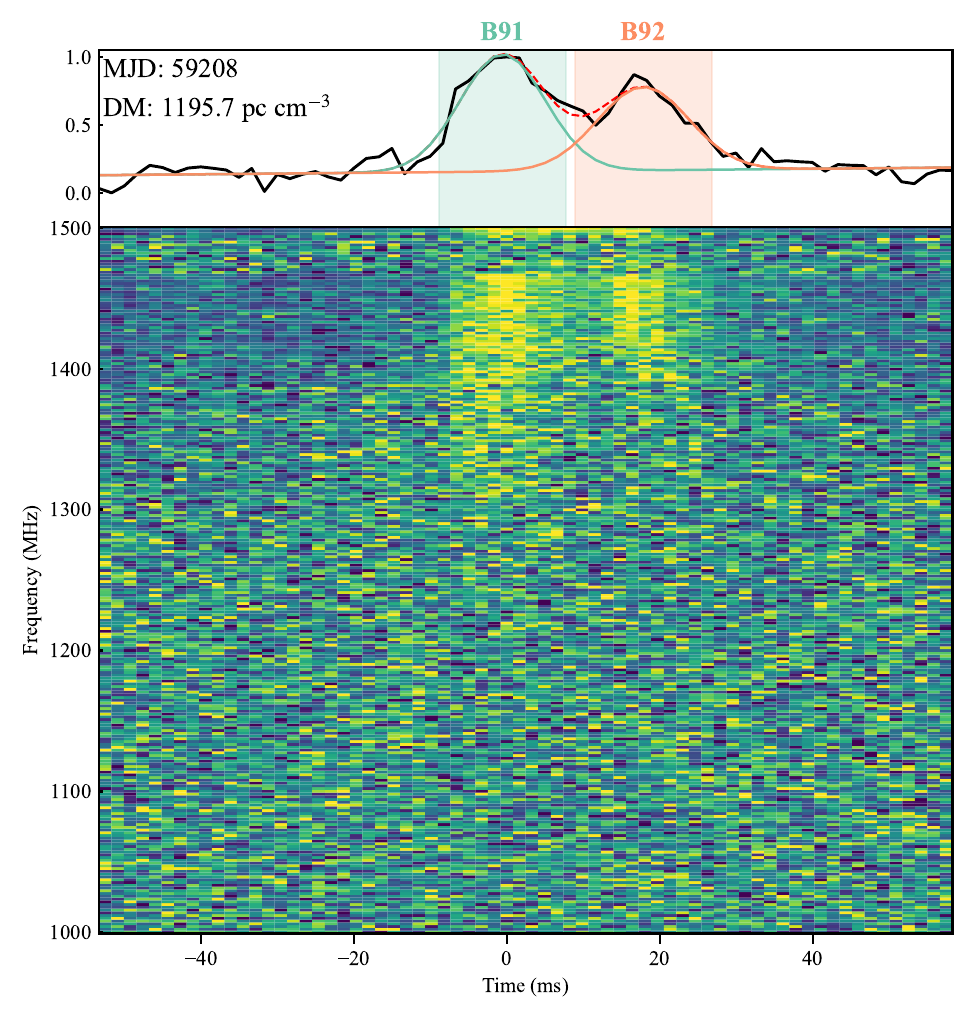}{0.29\textwidth}{}
        \fig{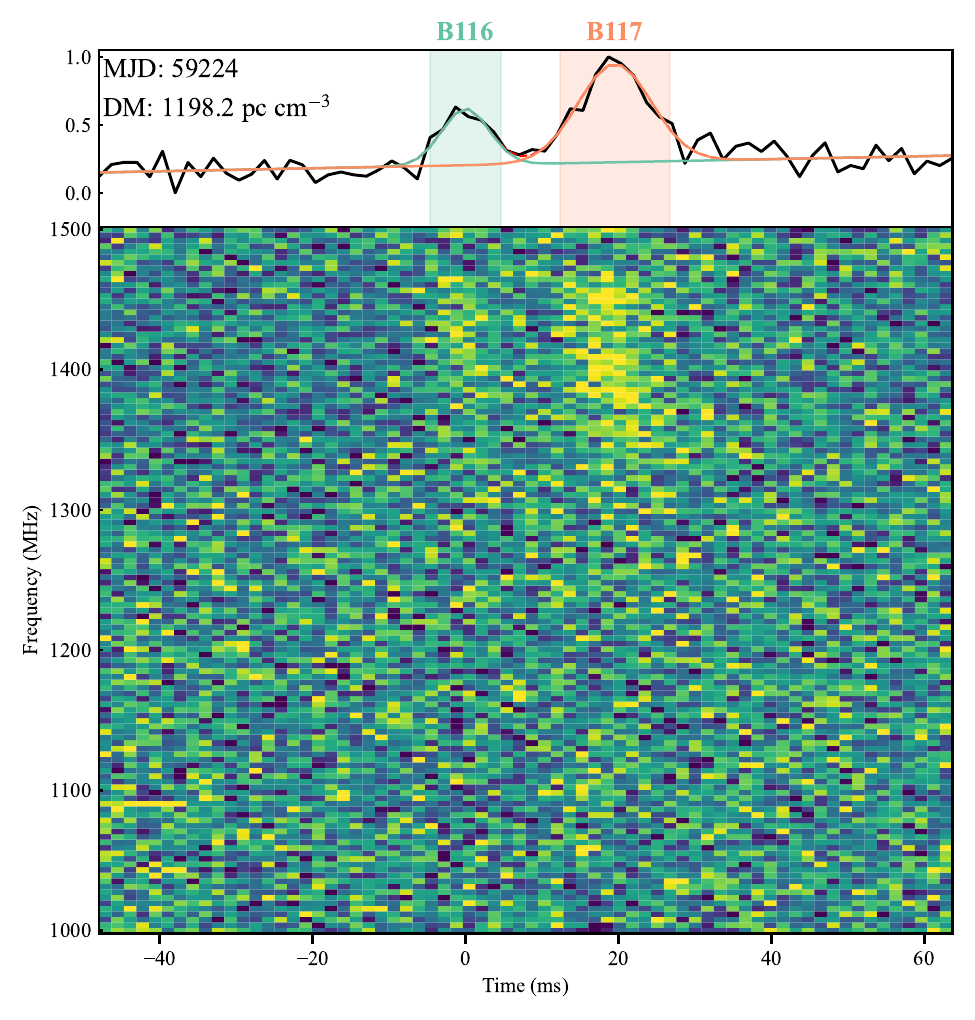}{0.29\textwidth}{}
        \fig{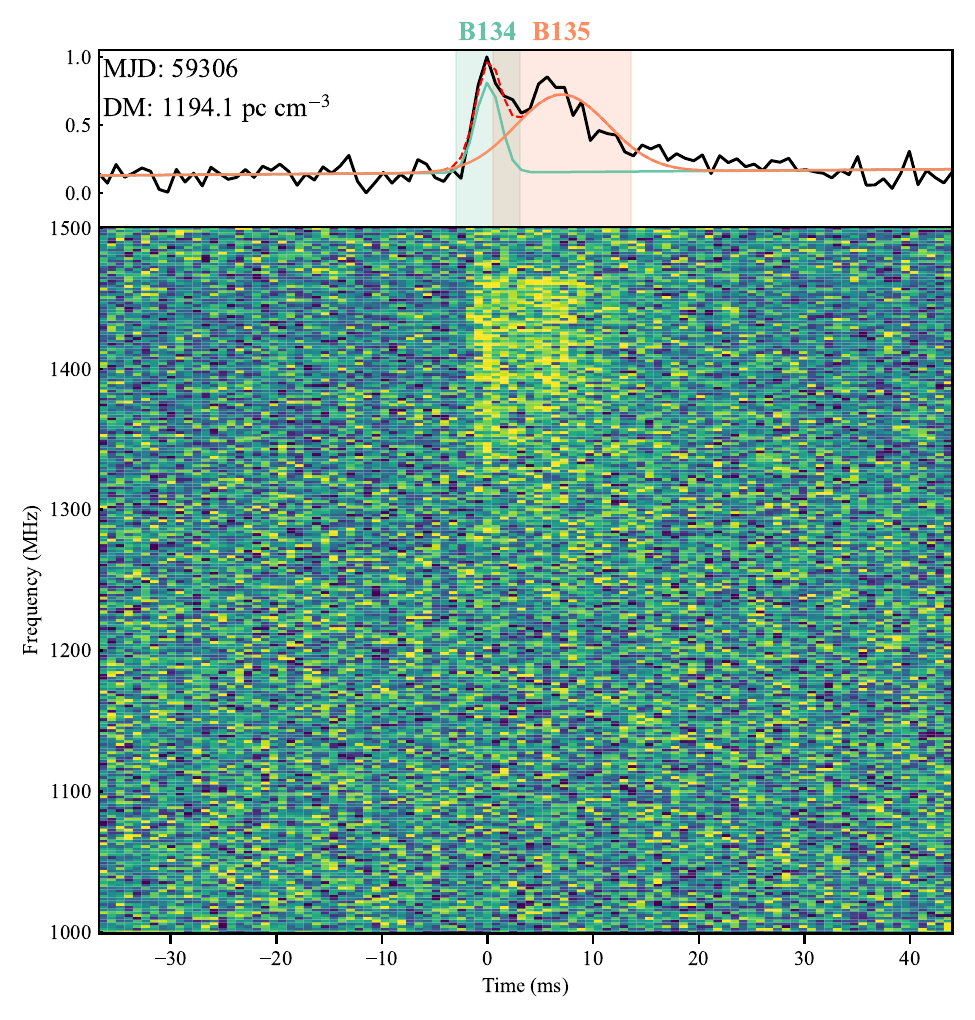}{0.29\textwidth}{}
    }
    \vspace{-0.9 cm}

    \gridline{
        \fig{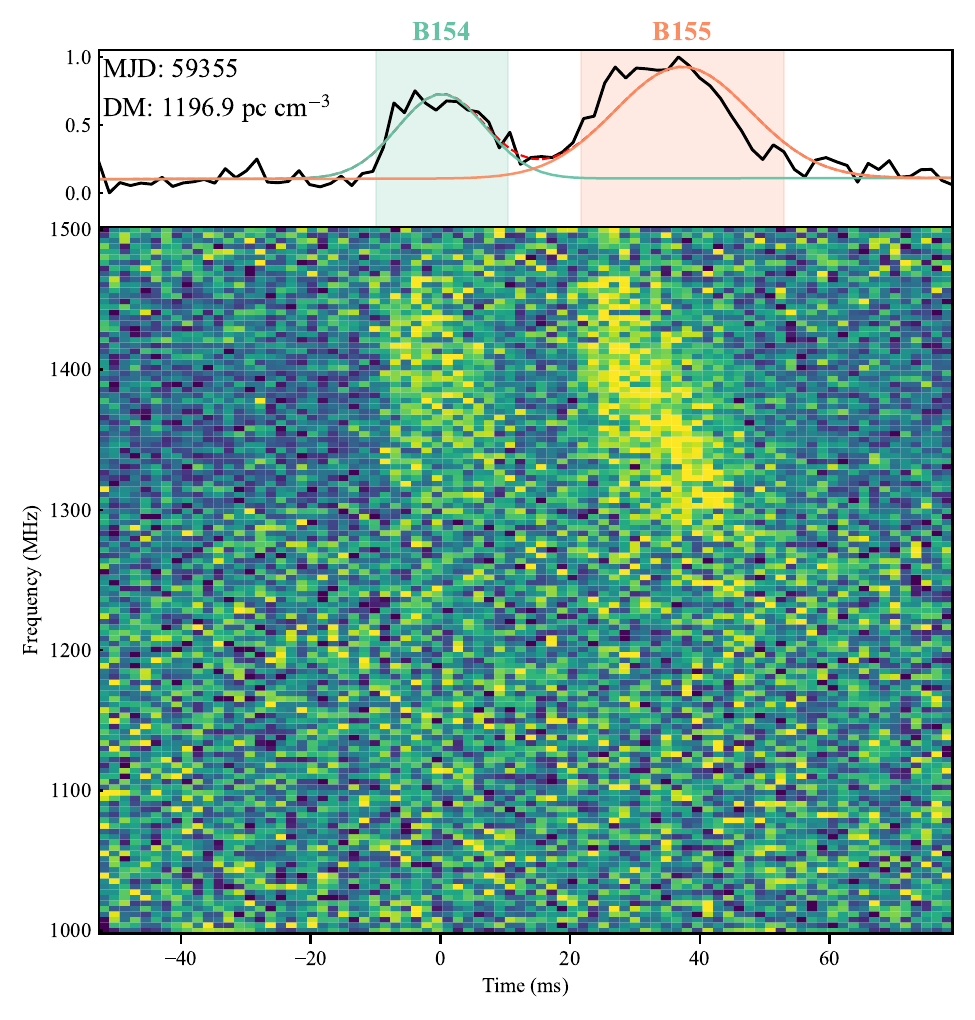}{0.29\textwidth}{}
        \fig{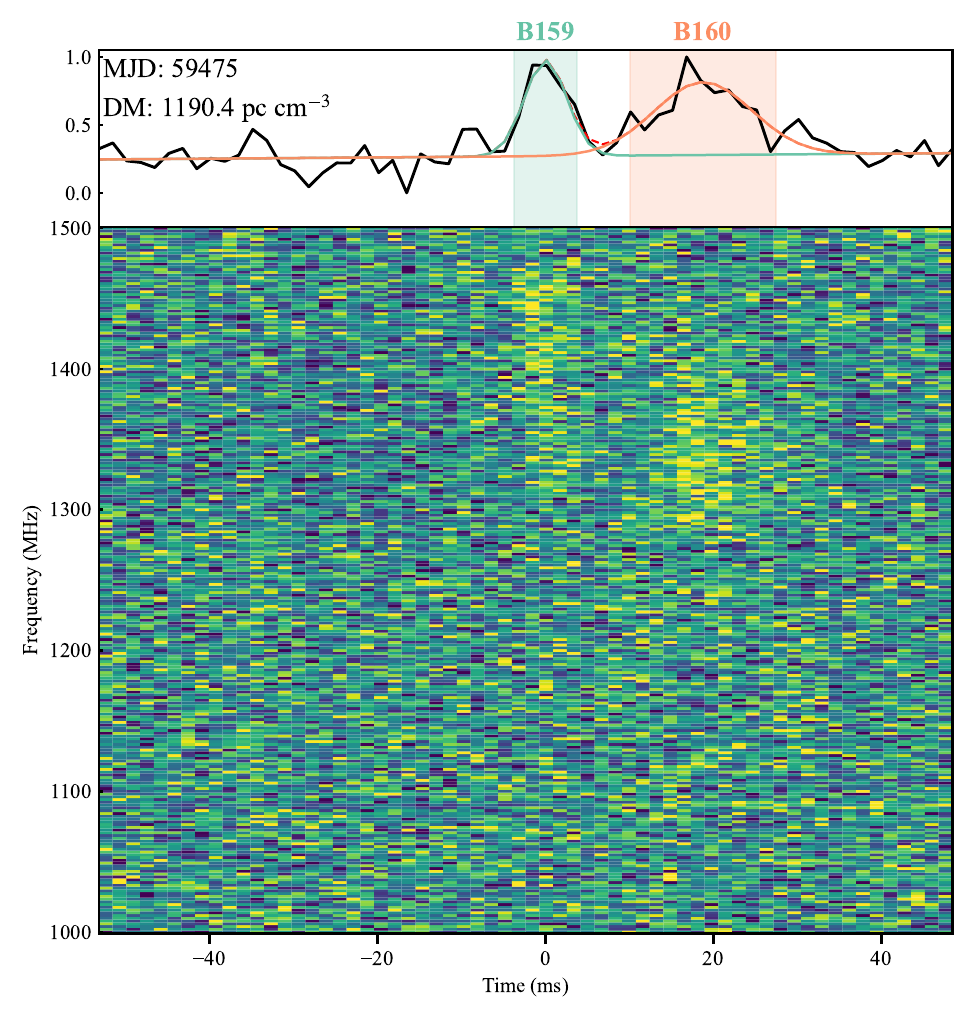}{0.29\textwidth}{}
        \fig{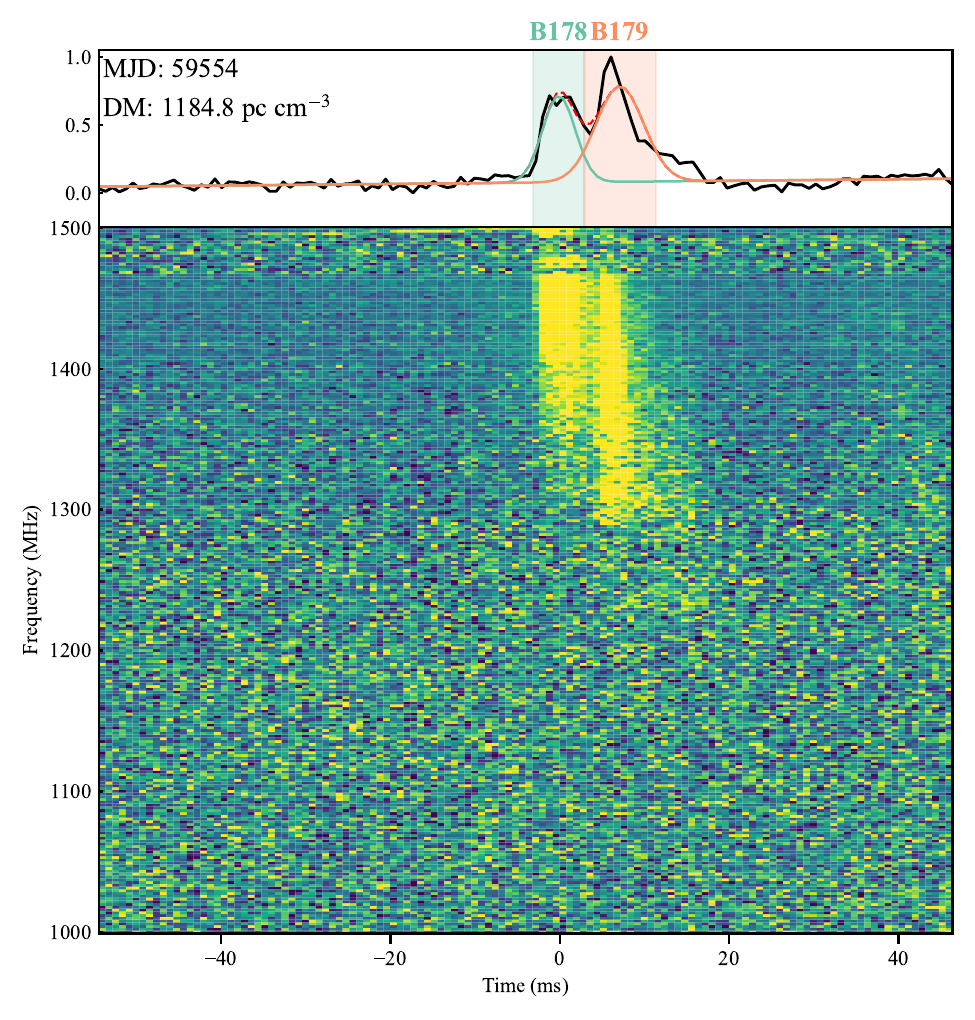}{0.29\textwidth}{}
    }
    \vspace{-0.9cm}

    \gridline{
        \fig{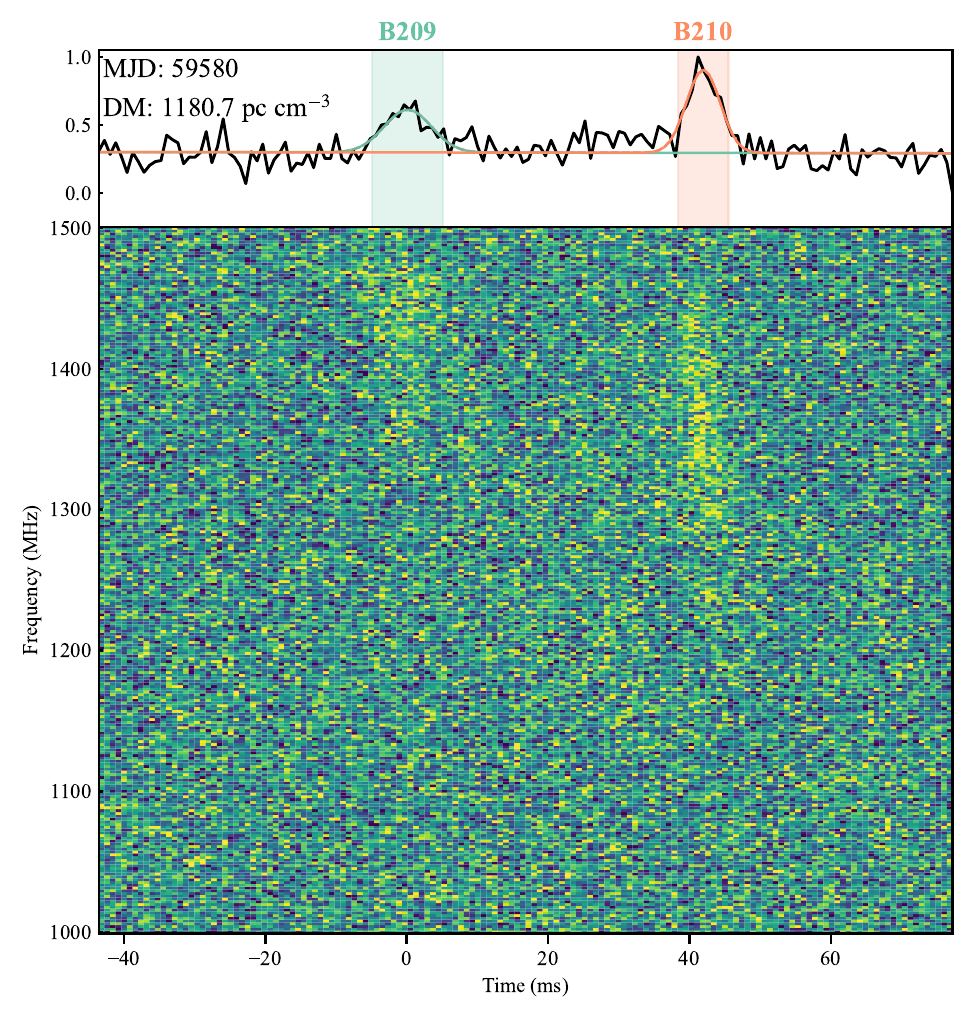}{0.29\textwidth}{}
        \fig{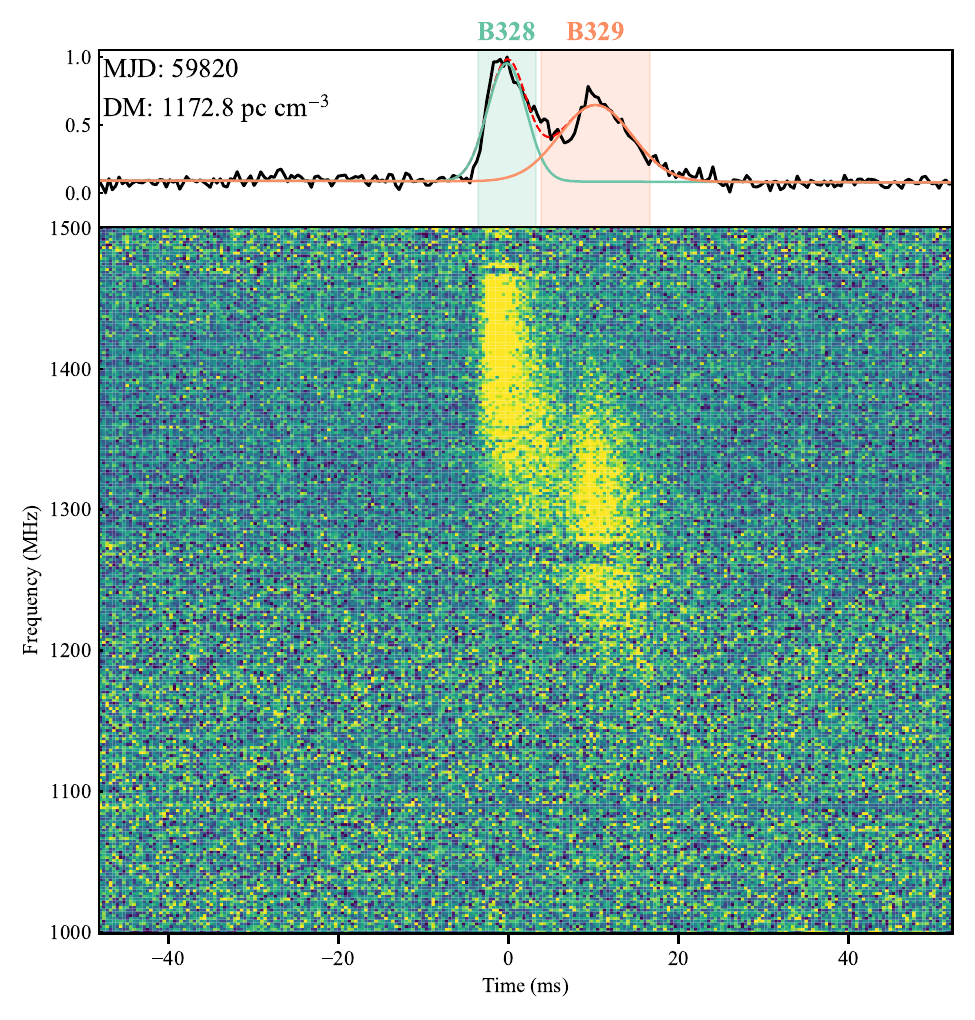}{0.29\textwidth}{}
        \fig{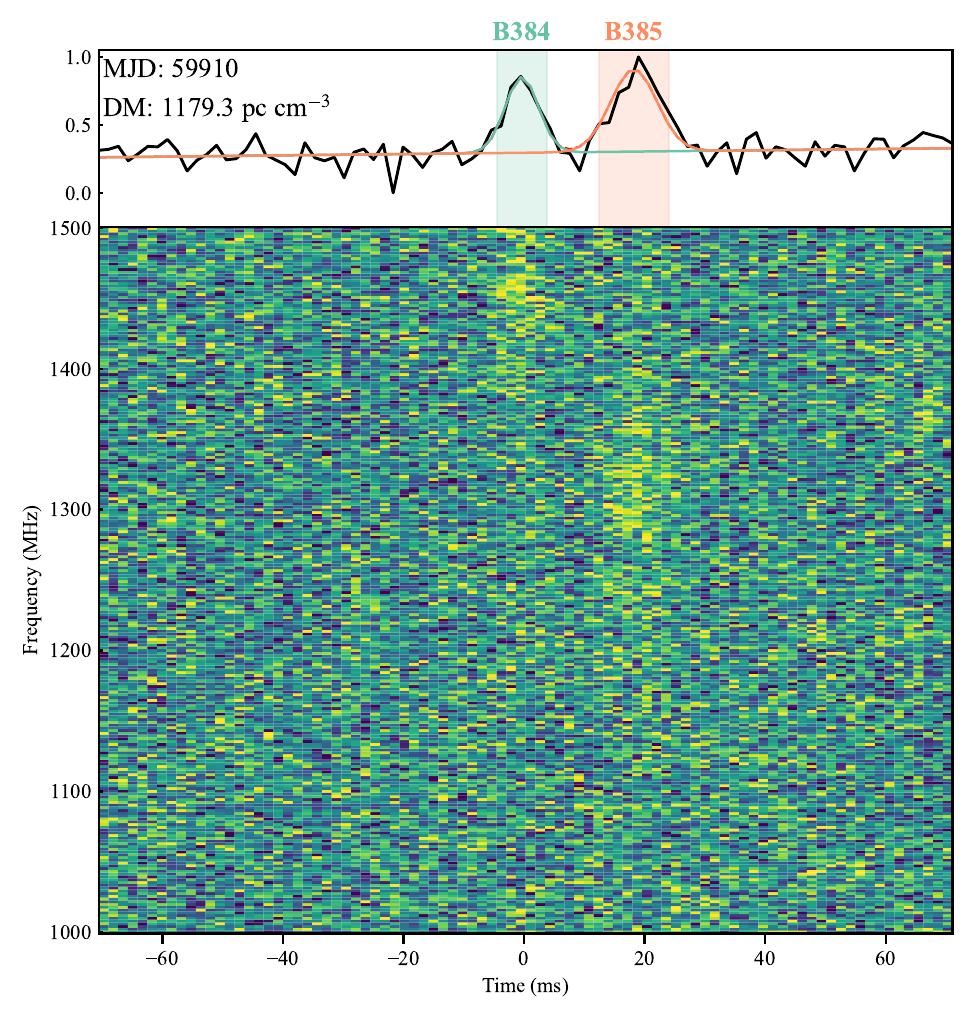}{0.29\textwidth}{}
    }
    \vspace{-0.9cm}
    
    \caption{\textbf{Two-dimensional dynamic spectra of multi-component burst-clusters from \frb.} 
    The top row presents a three-component burst-cluster, while the following three rows display nine representative two-component burst-clusters. 
    The three-component burst-cluster is labeled with indices 3-1, 3-2, and 3-3 beneath each panel. 
    The presentation follows the same format as Figure~\ref{fig:burst103}.}
    \label{fig:3-component burst}
\end{figure*}




\begin{figure*}[htbp]
    \centering
    \gridline{
        \fig{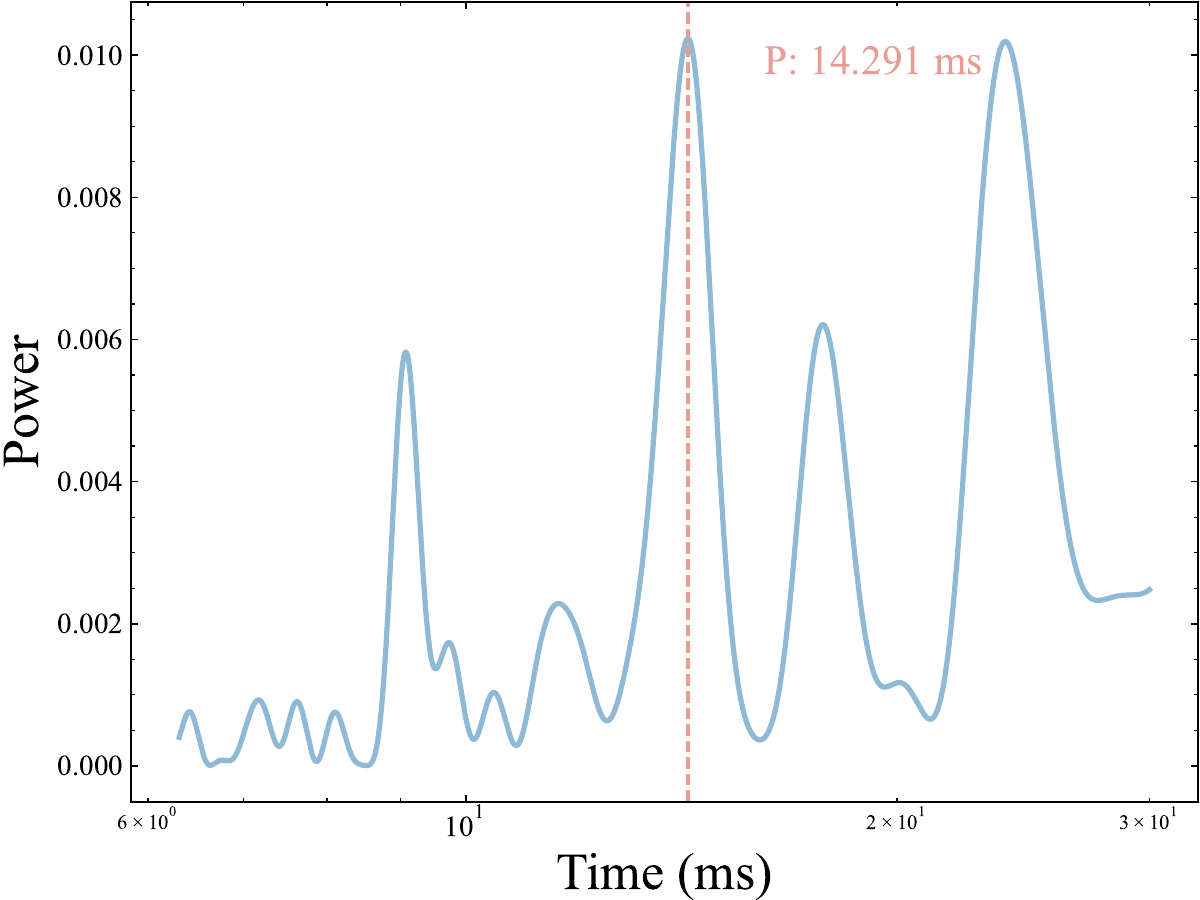}{0.24\textwidth}{3-1(a)}
        \fig{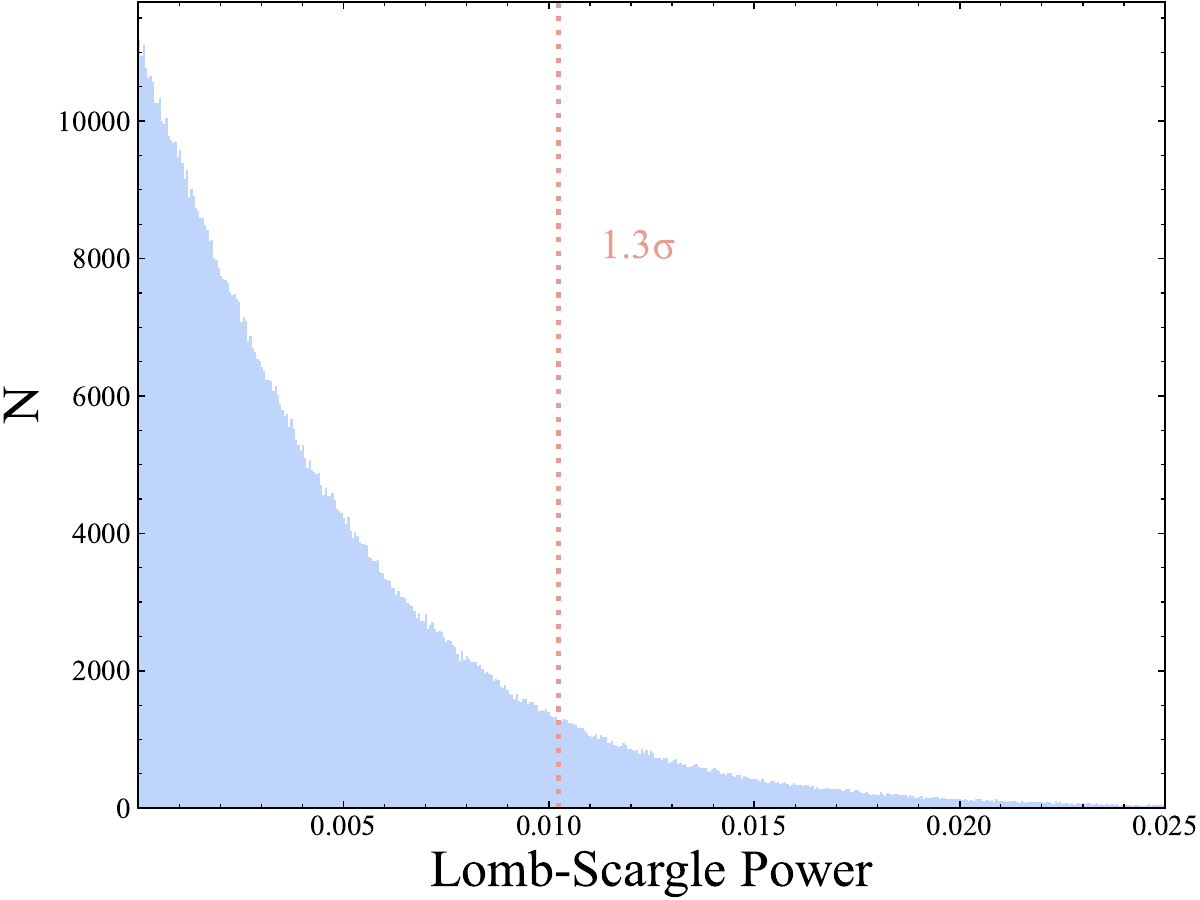}{0.24\textwidth}{3-1(b)}
        \fig{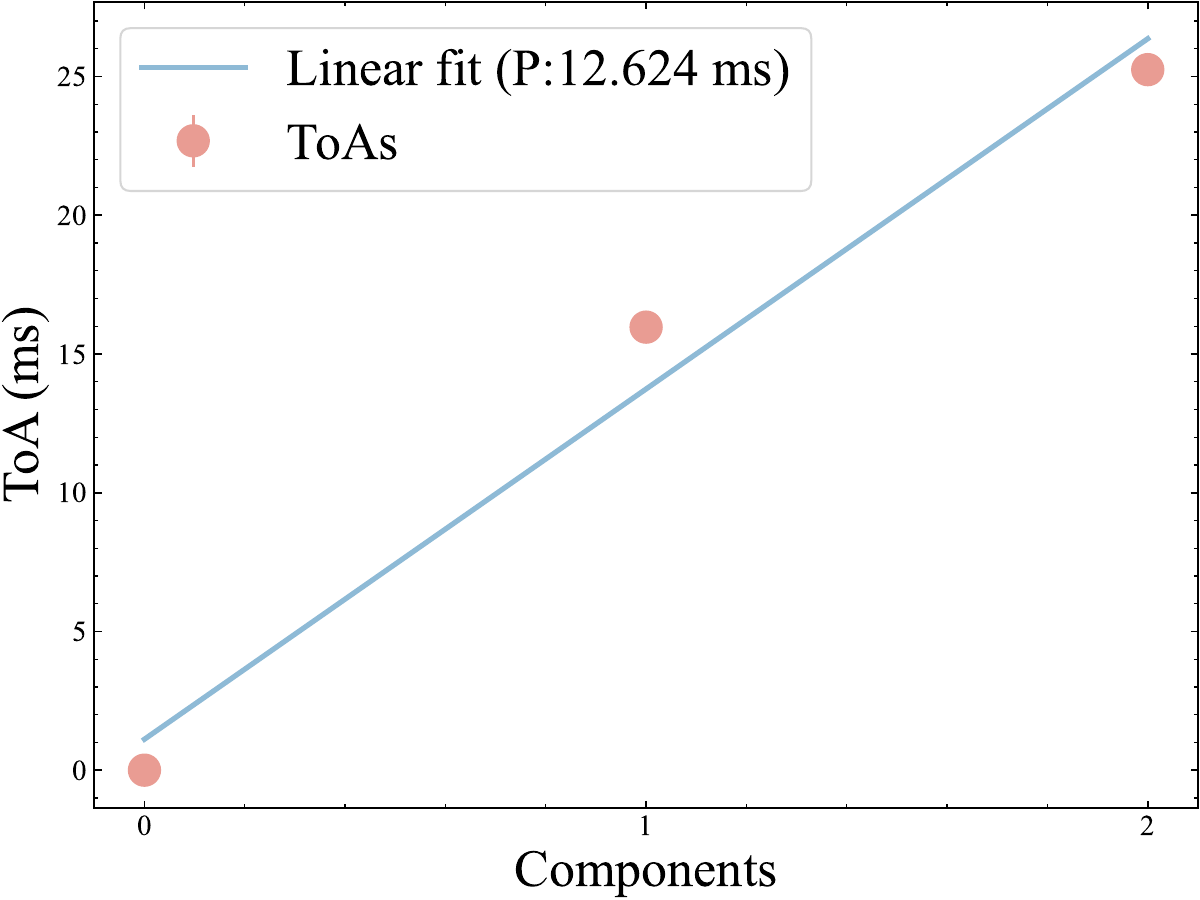}{0.24\textwidth}{3-1(c)}
        \fig{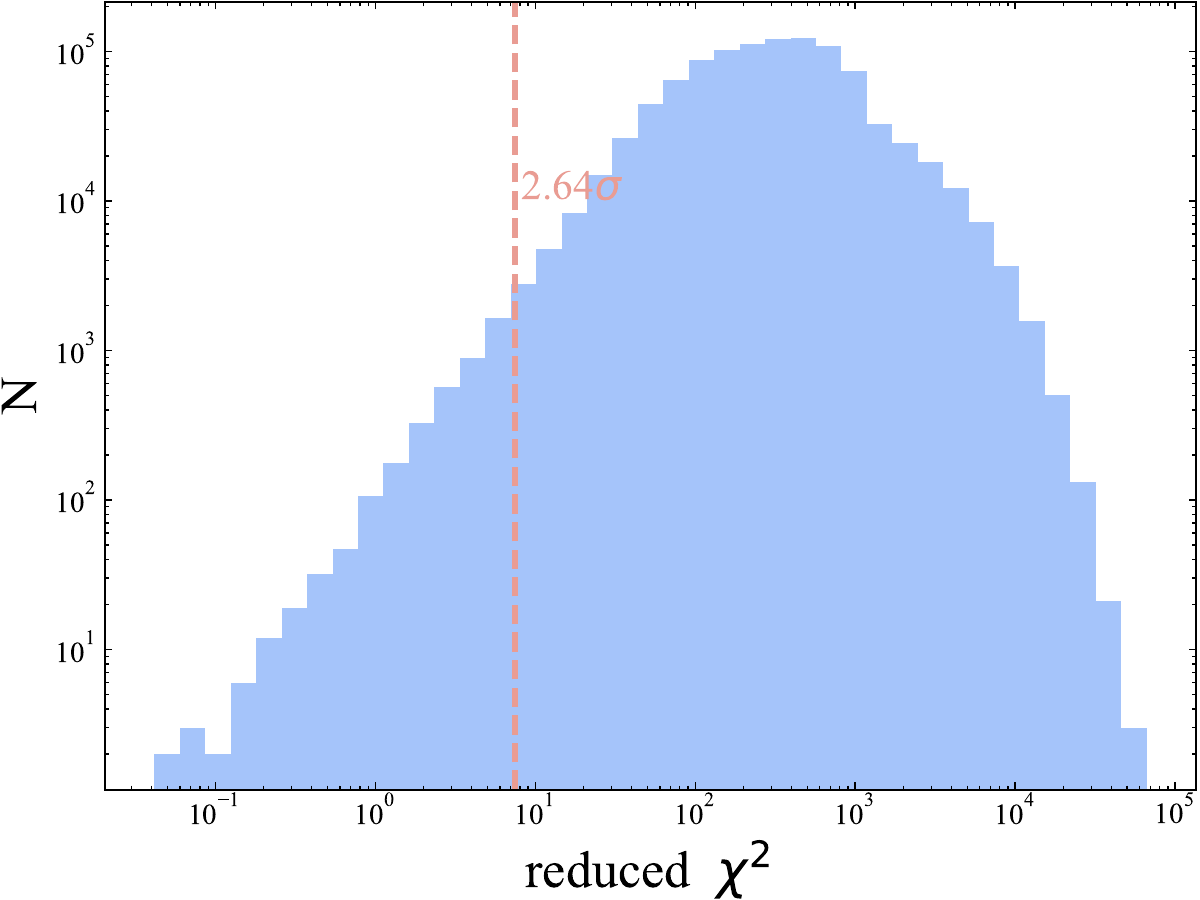}{0.24\textwidth}{3-1(d)}
    }
    \gridline{
        \fig{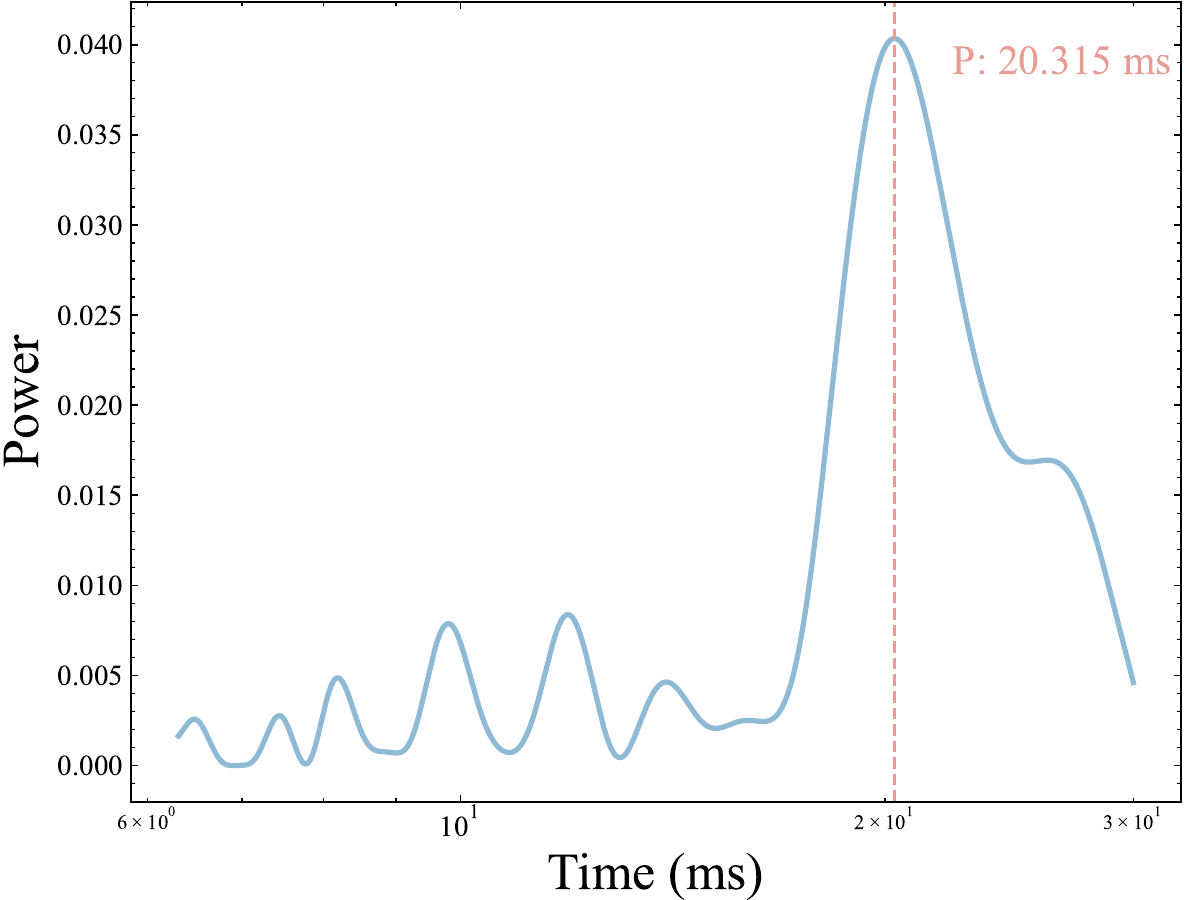}{0.24\textwidth}{3-2(a)}
        \fig{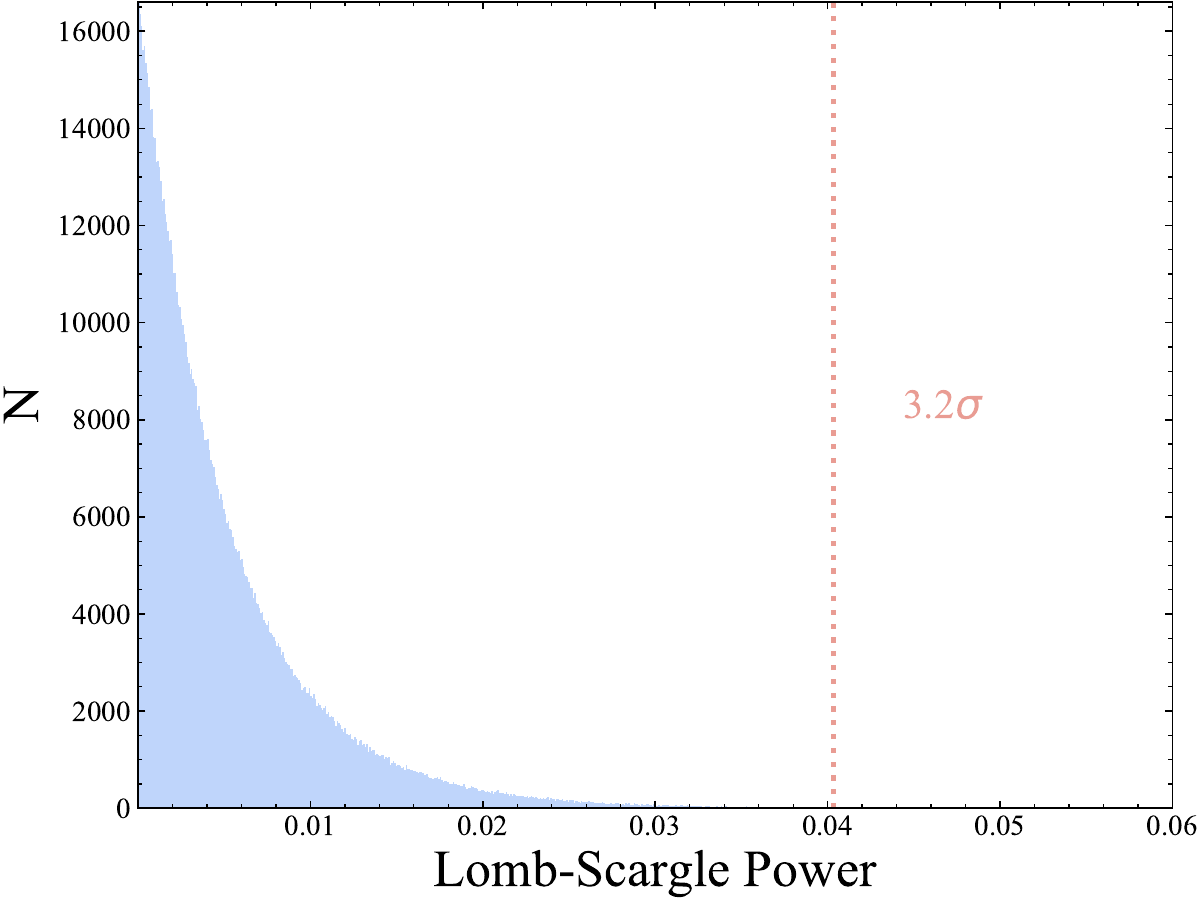}{0.24\textwidth}{3-2(b)}
        \fig{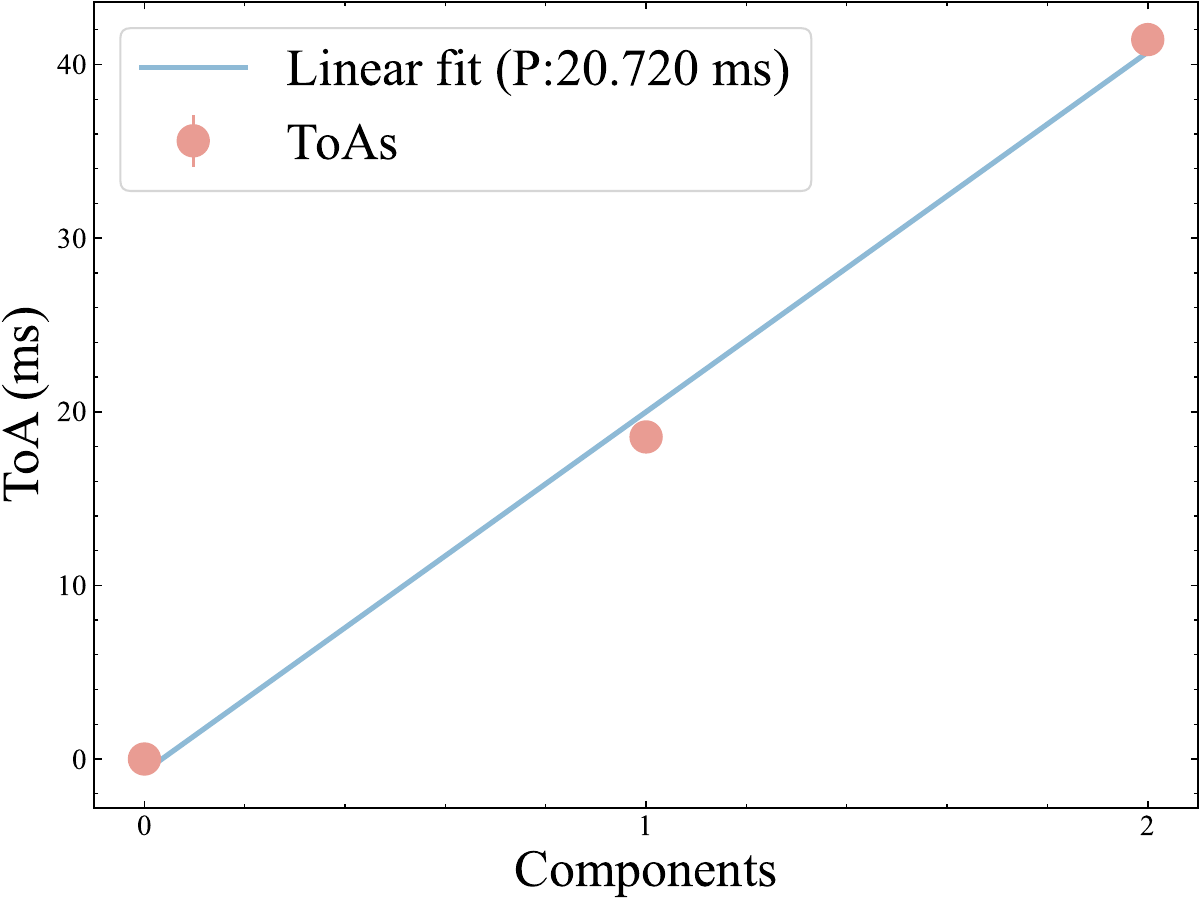}{0.24\textwidth}{3-2(c)}
        \fig{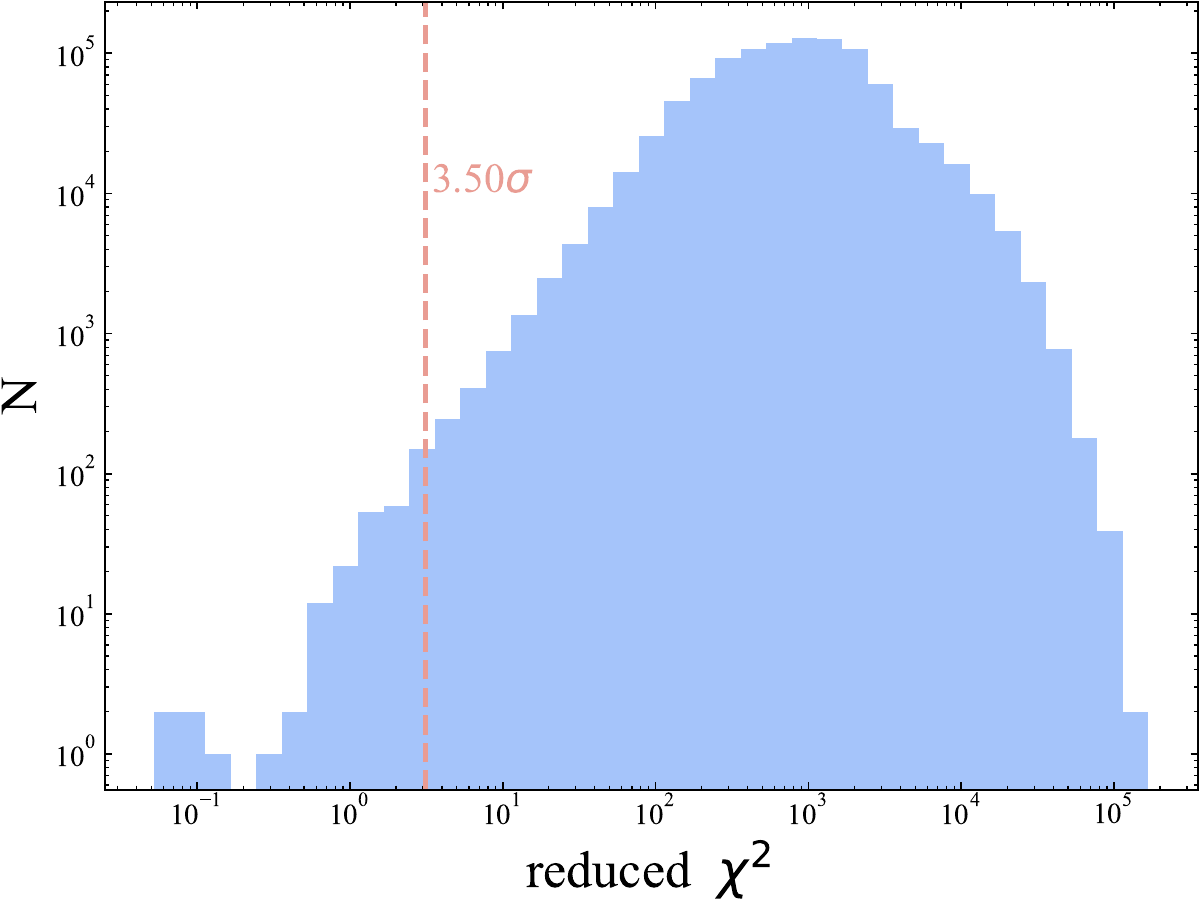}{0.24\textwidth}{3-2(d)}
    }
    \gridline{
        \fig{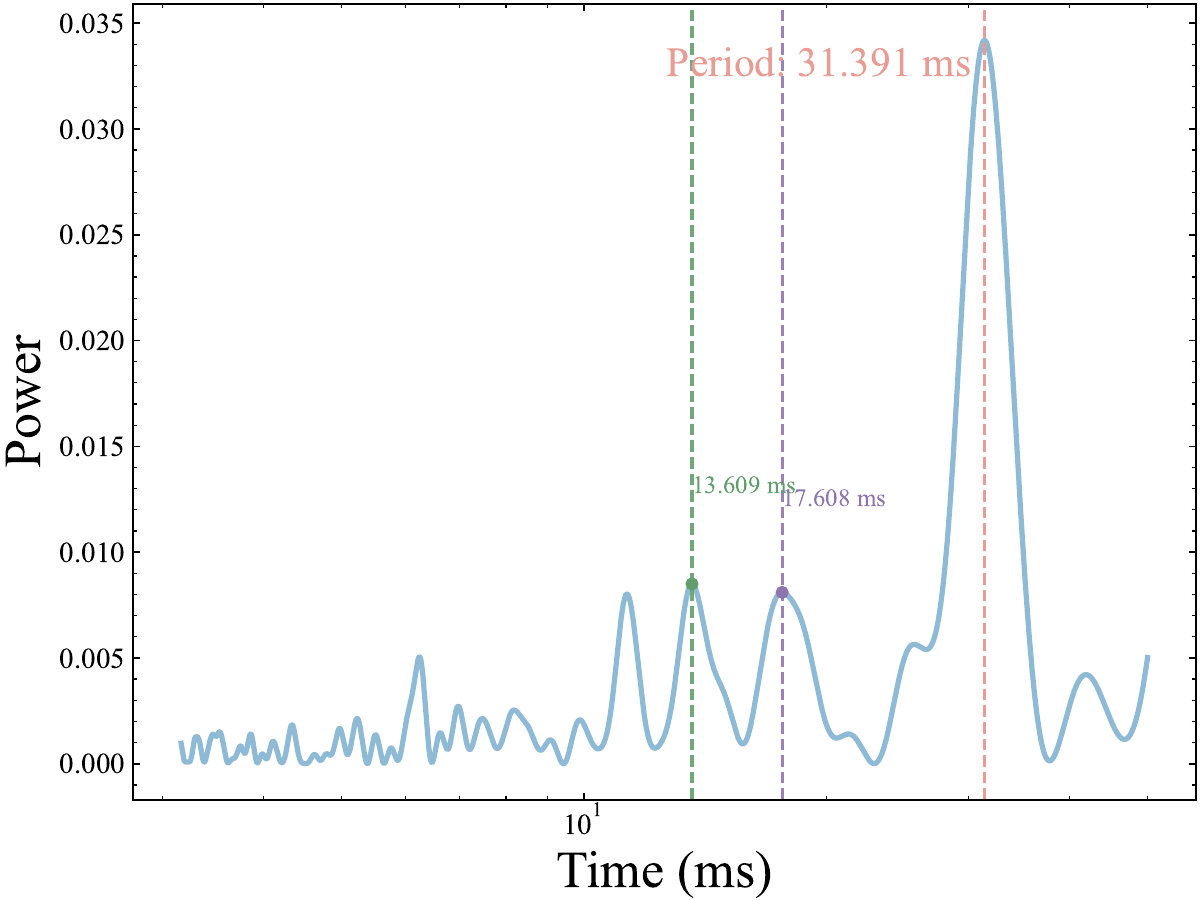}{0.24\textwidth}{3-3(a)}
        \fig{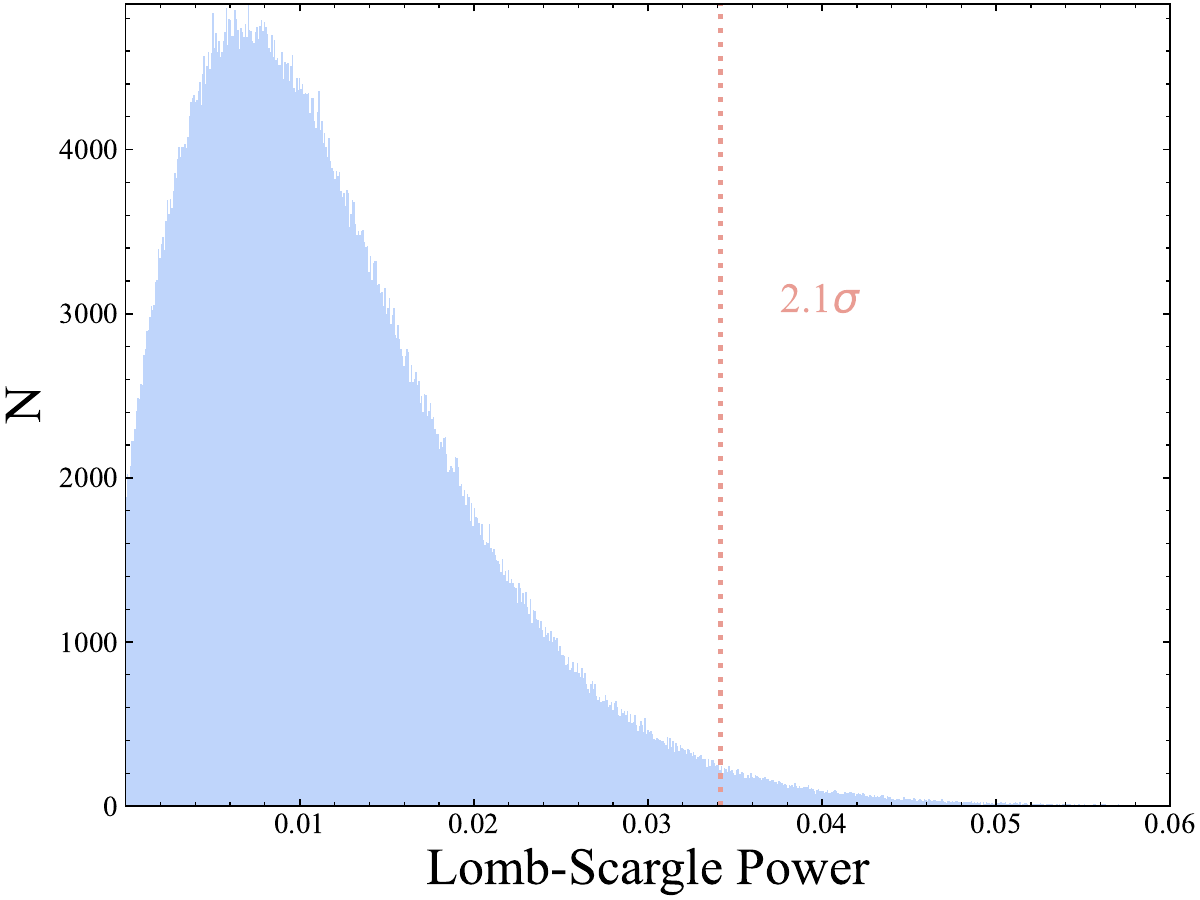}{0.24\textwidth}{3-3(b)}
        \fig{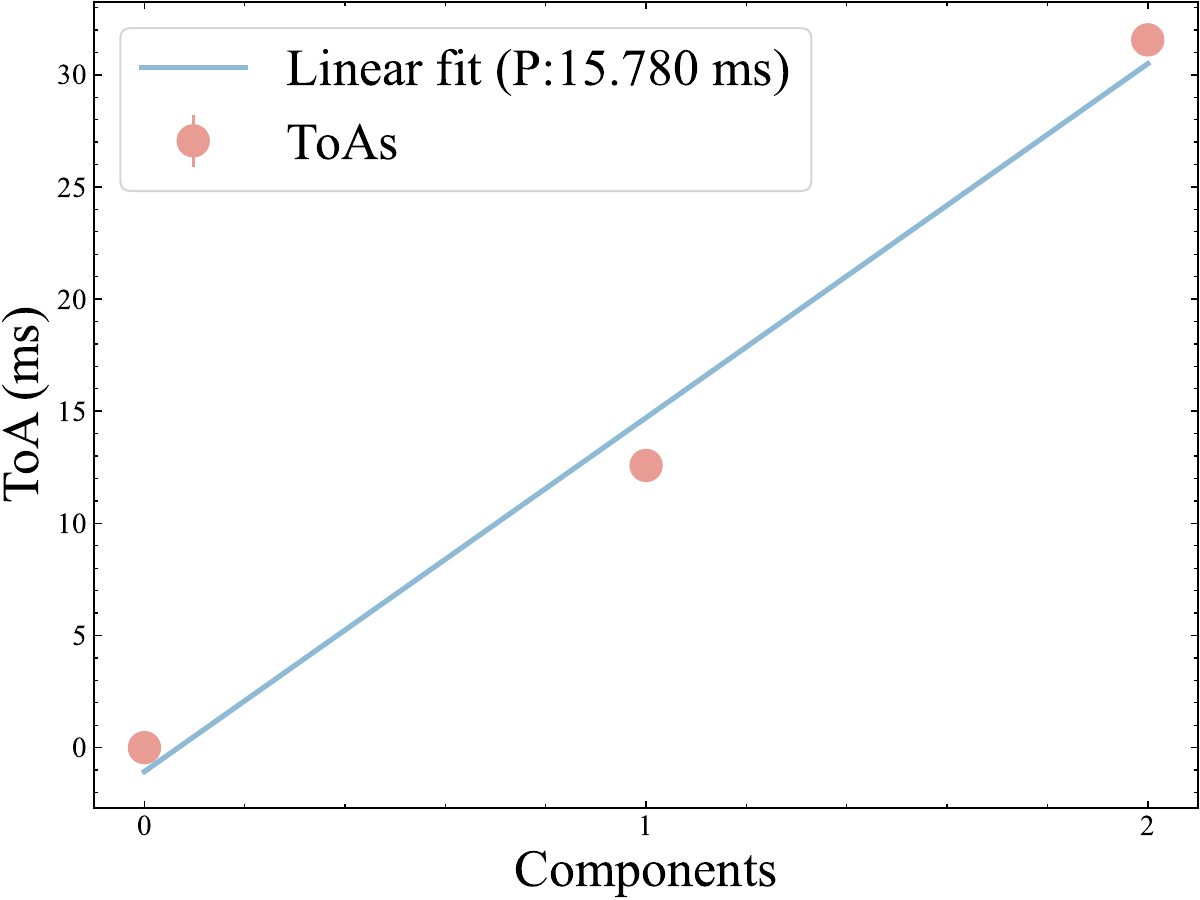}{0.24\textwidth}{3-3(c)}
        \fig{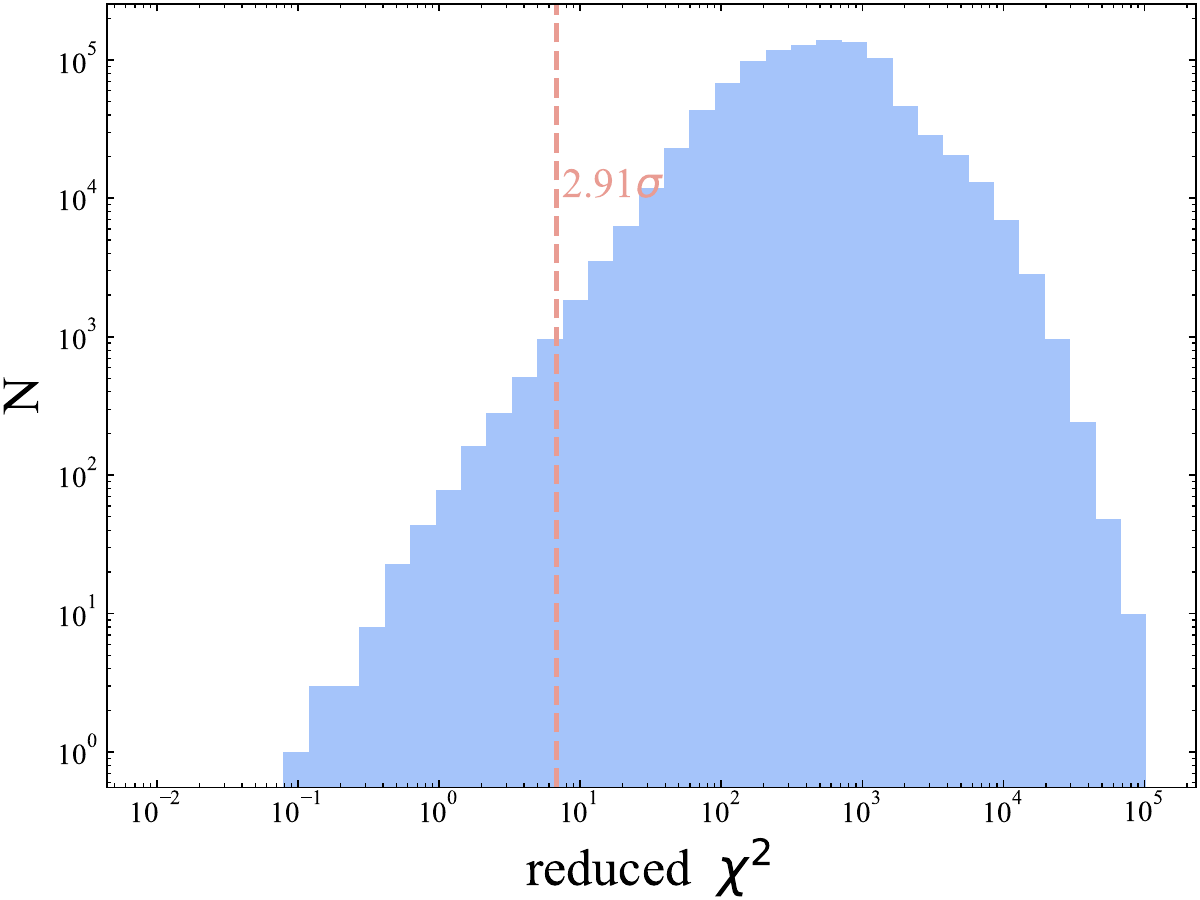}{0.24\textwidth}{3-3(d)}
    }
    \caption{Temporal analysis of the three-component burst-cluster. 
    Panels labeled as 3-1(a)–3-1(d), 3-2(a)–3-2(d), and 3-3(a)–3-3(d) correspond to the periodicity search results of the sub-panels 3-1, 3-2, and 3-3 in \rev{~\AppFig{fig:3-component burst}}, respectively. 
    The presentation follows the same format as Figure~\ref{fig:Temporal analysis}. 
    For the 3-3 case, panels (a) and (b) show distinct behavior: in panel (a), the highest Lomb–Scargle power corresponds to the time separation between the first and third peaks, rather than the expected adjacent spacing. This arises because the first and third pulses are broader in width compared to the second pulse. 
    The green and purple dashed lines mark the separations between peaks 1–2 and 2–3, respectively, for reference.}
    \label{fig:burst_series}
\end{figure*}

\end{document}